# A Dynamic Tree Algorithm for Peer-to-Peer Ride-sharing Matching

**Rui Yao**

Department of Civil and Environmental Engineering,

Technion – Israel Institute of Technology, Haifa 32000, Israel

Email: andyyao@campus.technion.ac.il

**Shlomo Bekhor**

Department of Civil and Environmental Engineering,

Technion – Israel Institute of Technology, Haifa 32000, Israel

Email: sbekhor@technion.ac.il

**ABSTRACT**

On-demand peer-to-peer ride-sharing services provide flexible mobility options, and are expected to alleviate congestion by sharing empty car seats. An efficient matching algorithm is essential to the success of a ride-sharing system. The matching problem is related to the well-known dial-a-ride problem, which also tries to find the optimal pickup and delivery sequence for a given set of passengers.

In this paper, we propose an efficient dynamic tree algorithm to solve the on-demand peer-to-peer ride-sharing matching problem. The dynamic tree algorithm benefits from given ride-sharing driver schedules, and provides satisfactory runtime performances. In addition, an efficient pre-processing procedure to select candidate passenger requests is proposed, which further improves the algorithm performance.

Numerical experiments conducted in a small network show that the dynamic tree algorithm reaches the same objective function values of the exact algorithm, but with shorter runtimes. Furthermore, the proposed method is applied to a larger size problem. Results show that the spatial distribution of ride-sharing participants influences the algorithm performance. Sensitivity analysis confirms that the most critical ride-sharing matching constraints are the excess travel times. The network analysis suggests that small vehicle capacities do not guarantee overall vehicle-kilometer travel savings.

**INTRODUCTION**

Traffic congestion is one of the major issues in many cities around the world. Not only it increases travel times and travel costs directly, but also imposes severe environmental problems to the public. There are several possible ways to alleviate congestion, and one of them is to increase vehicle occupancy. For example, the average vehicle occupancy rate is only around 1.5 in the USA (US National Household Travel Survey, 2017), which implies that many vehicles are traveling with empty seats. Since infrastructure expansion will not solve traffic congestions entirely, it is important to utilize the road infrastructure more efficiently.

Another way to alleviate congestion is to plan and operate public transport systems, which helps reducing private vehicle usage. Public transport vehicles typically take multiple passengers on board, and therefore can alleviate traffic congestions and vehicle emissions by reducing vehicle kilometers traveled (VKT). However, transit lines lack geographical and temporal flexibilities, because of their fixed routes and schedules. Moreover, it is costly and time consuming to increase public transit capacity and coverage. Although smarter, more efficient, and affordable new public transport options such as customized bus services (Qiu et al., 2014; Tong et al., 2017) are available, operators are still challenged to keep up with the demand. Taxi services, on the other hand, provide on-demand door-to-door transportation with high flexibility, but typically are more expensive and may worsen traffic congestion because of empty trips looking for passengers.

Emerging communication technologies and smartphone applications boost new alternatives for urban transportation. Innovative shared mobility services address the gap of conventional alternatives. They aim to provide more flexible mobility options than conventional public transits, and more economical alternatives than taxis, by reducing capital investments and operational costs.

One of the shared mobility alternatives is ride-sourcing. Service providers, such as Uber, Lyft, and DiDi, rely on smartphone applications to dispatch self-employed drivers. Essentially, ride-sourcing services are alternative taxi services with reduced fares, while online ride-sourcing companies use drivers' personal vehicles instead of owning vehicles. Passengers pay directly through the application to the drivers, and service providers typically keep a percentage of each fare and some other commission fees. However, there has been criticism that current ride-sourcing services are contributing to the growth of vehicles kilometer traveled in cities, because of empty car trips to pick up passengers and the profit-driven objectives of their matching algorithms (Clewlow and Mishra, 2017; Erhardt et al., 2019).

Therefore, another form of shared mobility, namely peer-to-peer ridesharing, aims to provide flexible on-demand mobility with similar quality of service as ride-sourcing, while able to address traffic congestions and reduce vehicle emissions. Peer-to-Peer ridesharing is based on the concept of shared economy (Lessig, 2008), in which collaborative consumption is split-up into single parts by the activities of sharing. In the context of peer-to-peer ridesharing, drivers are compensated for extra detours and to cover partially (or fully) their costs, the costs are then split between passengers who are provided with on-demand flexible transport with cheaper cost than taxi and reduce vehicle kilometer travel.

Different from ride-sourcing, peer-to-peer ride-sharing drivers have their designated destinations and want to arrive no later than some time. The trip purposes of peer-to-peer ride-sharing drivers are to perform activities other than only pick up and drop off passengers, i.e. they are not dedicated drivers. Peer-to-peer ride-sharing services are also provided by commercial companies, for example, Grab Hitch and DiDi Hitch. These services can be categorized into commercial, for-profit services, in which the commission fees are calculated based on the percentage of the fare. As a result, for-profit services may not maximize the potential of peer-to-peer ridesharing in reducing traffic and improving efficiency.

To fully utilize the potential of peer-to-peer ridesharing in terms of social objectives to reduce traffic congestions and environmental impacts, non-profit (or fixed-fee) peer-to-peer ride-sharing services can be provided by public (or private) entities to complement public transit by serving different trip types with private vehicles (Murphy, 2016; Hall et al., 2018), while being able to further reduce vehicle kilometer travel.

A variety of technical challenges are imposed on the implementation of on-demand peer-to-peer ride-sharing services. Specifically, the dynamic nature of peer-to-peer ridesharing makes the matching of drivers and passengers crucial to the success of the services. If we assume that drivers and passengers travel together between the same origins and destinations, the problem is relatively simple (Bahat and Bekhor, 2016). However, if passengers are scattered in the network the problem becomes particularly complicated. The matching algorithm needs to determine which passengers are served by the driver and a service sequence that satisfies different constraints also should be provided. This problem is related to the well-known dial-a-ride problem (DARP), which also tries to find the optimal pickup and delivery sequence for a given set of passengers.

This paper focuses on the peer-to-peer ride-sharing matching problem. A mathematical formulation of on-demand peer-to-peer ride-sharing matching problem based on a social objective to minimize the overall vehicle kilometer traveled is provided, and an efficient dynamic tree algorithm is catered for the on-demand ride-sharing matching problem. Given the relatively extensive literature on the subject, we first provide a concise literature review in the next section, which will help to pinpoint the contributions of the present paper.

**Literature Review**

The ride-sharing matching problem can be viewed as a generalization of the DARP which is well studied in the literature. Psaraftis (1980) modeled the DARP with a single vehicle under static and dynamic settings. Savelsbergh and Sol (1995) formulated the basic form of DARP, where all vehicles depart and arrive at the same depot. In a more realistic DARP setting, time window constraints for the passenger requests (Psaraftis, 1983), multiple depots for the vehicles (Cordeau and Laporte, 2007), and heterogeneity of vehicles capacities and number of passengers in a request (Braekers, Caris and Janssens, 2014) are included. Note that despite the similarities between DARP and ride-sharing matching problems, there exists a key difference between on-demand peer-to-peer ride-sharing services and dial-a-ride services: instead of being employees of the system in a dial-a-ride service, ride-sharing drivers are considered as clients who have their distinct origins, destinations, and time constraints.

Different approaches are developed in the literature to solve DARP. Cordeau (2006) introduced branch and cut (B&C) algorithm by applying several families of valid inequalities as cuts. Ropke et al. (2007) also introduced reachability constraints from the vehicle routing problem (VRP) with time windows to the B&C algorithm. Other heuristics and metaheuristics were further developed to handle large-size DARP. For example, simple insertion heuristics were proposed to quickly find feasible solutions where each request is inserted in the vehicle's schedule by the cheapest insertion criterion (Jaw et al., 1986; Xiang et al., 2008; Wong et al., 2014). Cordeau and Laporte (2003) introduced Tabu search by locating requests in different neighbors with additional heuristic diversification strategies.

Ride-sharing systems are presented in different forms. In the simplest form of ride-sharing, the system will try to match each driver with one passenger only (Agatz et al., 2011). Yan et al. (2019) also studied the single passenger ride-sharing stochastic at a macroscopic level with user equilibrium. In order to fully utilize the unused capacity of the vehicles, more sophisticated ride-sharing systems allow multiple passengers assigned to a single vehicle (Herbawi and Weber, 2012; Alonso-Mora et al., 2017; Simonetto el al., 2019) or even allow passengers transfer between vehicles (Herbawi and Weber, 2011; Herbawi and Weber, 2012; Masoud and Jayakrishnan, 2017), solving these forms of ride-sharing matching problem may require exploration of large decision space.

In this paper, we focus on the multi-passenger ride-sharing matching problem without the need for the passenger to transfer between vehicles. The ride-sharing matching problem contains mainly the passenger-driver assignment problem and the VRP/DARP. Because of the similarity between DARP and ride-sharing matching problem, solution algorithms for DARP can be applicable to the ride-sharing matching problem. More efficient algorithms accounting for the dynamic nature of ride-sharing matching problem were also developed in the literature.

In terms of algorithmic implementations, solution approaches can be categorized into one-to-many algorithms and many-to-many algorithms. One-to-many algorithms are passenger-oriented, since they try to find the best vehicle with respect to the objectives for a given passenger, and adjust the route to serve the given passenger. Many-to-many matching algorithms consider a group of drivers and passengers at the same time, in which multiple passengers are matched with drivers while satisfying time window and capacity constraints. Since many-to-many matching algorithms optimize the global matching between drivers and passengers for a period of time, they have the potential to achieve better matches than one-to-many algorithms, which only optimize the local matching for a single passenger. However, since many-to-many algorithms are intrinsically more complex, efficient approaches to solve the problem still require further investigations.

There are several one-to-many algorithms developed in the literature. The approach proposed by Huang et al. (2017) calculates the costs of all feasible permutations and finds the exact least-cost feasible vehicle schedule. Moreover, a dynamic tree structure that maintains only valid vehicle schedules were implemented. In this way, only the feasible subset of all possible permutations needs to be considered. Different heuristics were proposed to speed up the ride-sharing matching process. For example, Ma et al. (2015) suggested nearest neighborhood vehicle search heuristic based on spatial indexing, and Jung et al. (2016) proposed a hybrid-simulated annealing method where Euclidean distance is used as an insertion heuristic to maintain the solution feasibility.

Different approaches were proposed to solve the ride-sharing matching problem using many-to-many algorithms. Early works were only able to solve small size problems in a reasonable amount of time. Baldacci et al. (2004) formulated the problem into two integer linear programs and proposed a bounding heuristic to solve the problem. Herbawi and Weber (2012) solve the matching problem using genetic algorithm. Santos and Xavier (2013) applied greedy randomized adaptive search in the ride-sharing matching problem by adaptively constructing diverse initial solutions.

Recent works have developed more efficient algorithms in the context of real-time applications. In Agatz et al. (2011), the single-passenger ride-sharing matching problem is formulated and solved by maximum weighted bipartite matching. To reduce the search space, Masoud and Jayakrishnan (2017) introduced the ellipsoid spatiotemporal accessibility method (ESTAM) to construct the passenger's time expanded feasible network and solved the matching problem using dynamic programing; Najmi et al. (2017) proposed a clustering heuristic based on Euclidean distance to reduce the problem size. In Simonetto et al. (2019), the ride-sharing matching problem is simplified to a single-passenger assignment problem and solved using linear programming. Alonso-Mora et al. (2017) proposed a general many-to-many dynamic multi-passenger vehicle assignment framework by reducing the problem to a passenger-combination (or trip) to driver matching problem. Their approach decouples the problem by first checking the shareability of passengers and drivers based on the idea of shareability networks (Santi et al., 2014). The algorithm starts by constructing a pairwise passenger-driver graph in which passengers connect to the same driver have the potential to share the ride; second, feasible passenger combinations are generated for each driver; third, an integer linear program (ILP) is solved to match passenger-combination to drivers.

Recent studies extended the matching problem and corresponding solution algorithms. For example, investigating the stability of the matching between passengers and drivers (Wang et al., 2018), improving the traffic efficiency with advanced travel time feedback (Wu et al., 2019), and optimizing the many-to-many matching time interval and matching radius (Yang et al., 2020). Studies concerning the effects of transportation services on emergency events, such as epidemic disease spreading (Chen et al., 2020a, 2020b) were also recently investigated in the literature.

**Contribution**

In this paper, we consider a flexible setting, in which drivers are willing to make a detour to cover both the pickup and drop-off locations of the passengers. Potential ride-sharing participants announce their trips to the system as a passenger or a driver at a time close to their desired departure time. A trip announcement will include the origin-destination locations and its specific time windows. With this given information, the ride-sharing system will try to assign multiple passengers to each vehicle, and determine the detour to pick up and drop off the passengers.

Yao, Bekhor

Our proposed dynamic tree algorithm falls in the category of many-to-many algorithms, in which multiple passengers are considered and assigned to drivers at the same time. Our proposed framework includes a preprocessing procedure and an efficient ride-sharing VRP algorithm which makes our algorithm more efficient.

The contributions of the present paper are outlined as follows:

1. We propose a geometric pruning procedure to efficiently eliminate the infeasible passenger requests for each driver based on the spatiotemporal proximity of both drivers and passengers. This approach is able to accommodate the multi-passenger multi-driver problem with the introduction of reachable pickup region of the passenger as a circle. Candidate passengers are filtered by finding the intersection between reachable pickup regions and accessible regions of drivers.

2. We adapt and integrate a dynamic tree structure to solve the ride-sharing VRP. Within the many-to-many framework, our proposed dynamic tree algorithm is able to efficiently find high-quality matches for all drivers and passengers, instead of local matches for each passenger. Moreover, our algorithm is more general than Huang et al. (2017) in the sense that any number of passengers can be assigned to a vehicle (provided there are available seats). Additional passengers can be added to the vehicle that currently has no empty seat after a drop-off occurs and further utilize the vehicle capacity.

3. The dynamic tree algorithm utilizes previous routing computations, and significantly improves computation times. Results obtained using simple networks show that the dynamic tree algorithm reaches the same objective function values of the exact algorithm, but with significantly shorter runtimes.

4. In addition to the algorithmic contributions, we perform simulation studies to assess the impact of ride-sharing services on the overall transportation network performance, using the well-known Winnipeg network.

**PROBLEM FORMULATION**

In a peer-to-peer ride-sharing setting, new trip announcements (of both drivers and passengers) arrive in the system. In the context of many-to-many matching algorithms, trip announcements are collected and solved in a batch. For simplicity, we assume a relatively large time period (e.g. 1 hour), in which requests arrive during this time period are collected into a batch and solved together. The performance of the on-demand peer-to-peer ride-sharing system is highly dependent on the algorithm for solving this problem; therefore, our paper will focus on determining the matches for a batch of requests.

**Passenger-Driver Network**

In order to define the problem (and later improve the efficiency of the algorithm), we will first construct a passenger-driver network $G(N, A)$, in which each passenger or driver is associated with a distinct origin and destination node. We define the set of nodes for origins and destination of drivers $V$ as $N_V$, set of nodes for pickup and drop-off locations of passengers $R$ as $N_R$. In cases that multiple participants share the same physical node in the road network, each participant is assigned a distinct duplicated node in the passenger-driver network $G$. Each arc $(a, b) \in A$ represents a shortest path from node $a$ to node $b$ with travel time on the arc as $tt_{(a,b)}$ and distance as $l_{(a,b)}$.

Yao, Bekhor

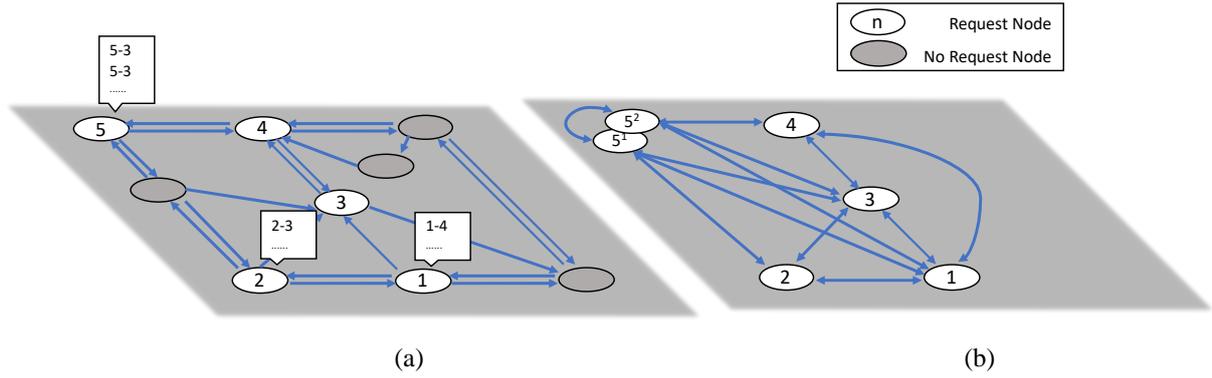

Figure 1 Physical road network (a) and Passenger-driver network (b)

In Figure 1, we illustrate a physical road network (a) and the corresponding passenger-driver network (b). For example, node 5 in the physical network is associated with 2 requests, this node is duplicated into 2 nodes as $5^1$ and $5^2$. In case that there are feasible paths between every passenger request and driver node, the corresponding network is represented by a complete graph (that is, a fully connected graph).

**Ride-Sharing Matching Problem**

We consider a ride-sharing setting that participant $i$ specifies his/her role as a driver or a passenger, respectively denoted by $v_i$ or $r_i$. The set of ride-sharing drivers is $V = \{v_1, \dots, v_n\}$ and the set of passenger requests is $R = \{r_1, \dots, r_n\}$. Participant $i$ needs to provide information about his/her origin $o_i$ and destination $d_i$, maximum excess travel time $\Delta_i$ compared to his/her shortest path travel time $\tau(o_i, d_i)$, and his/her earliest departure time $t_{ED}(v_i)$ or $t_{ED}(r_i)$ for driver and passenger respectively. For ride-sharing driver $v_i$, he/she also needs to provide the number of passengers he/she is willing to take (or vehicle capacity), defined as $c_i$. For ride-sharing passengers, the maximum waiting time for pickup is defined as $\Omega_i$.

Our ride-sharing formulation tries to optimally assign the passenger requests to the vehicles, and finds the optimal routes for the vehicles. The optimal solution is of course dependent on the objective function sought, which is discussed in the following paragraphs.

We define the decision variable as $x^{v_i}_{(a,b)} \in \{0,1\}$; if the route of driver $v_i$ passes through arc $(a, b)$ in the passenger-driver network, then $x^{v_i}_{(a,b)} = 1$, which indicates driver $v_i$ is assigned to either pickup or drop off passengers at node $a$ and $b$. The matching between drivers and passengers may increase the travel time for the ride-sharing driver. Thus, the excess travel time $\delta_{v_i}$ for ride-sharing driver $v_i$ is defined as:

$$\delta_{v_i} = \sum_{(a,b) \in A} tt_{(a,b)} \cdot x^{v_i}_{(a,b)} - \tau(o_i, d_i) \tag{1}$$

where the first part represents the driver's actual travel time, and $\tau(o_i, d_i)$ is the shortest travel time between $o_i$ and $d_i$.

The waiting time $\omega_{r_j}$ of passenger $r_j$ assigned to driver $v_i$ is defined as:

$$\omega_{r_j} = \sum_{v_i \in V} \left( t_{v_i}(o_j) - t_{ED}(r_j) \right) \cdot z^{v_i}_{r_j} \tag{2}$$

Where $t_{v_i}(a)$ represents the actual arrival time of driver $v_i$ at node $a$, the auxiliary variable $z^{v_i}_{r_j} \in \{0,1\}$ represents a passenger request $r_j$ assigned to driver $v_i$. Note that this variable is dependent on $x^{v_i}_{(a,b)}$. The difference between the actual pickup time and the earliest pickup time is the waiting time.

The excess travel time $\delta_{r_j}$ of passenger $r_j$ is defined as:

$$\delta_{r_j} = \sum_{v_i \in V} \left( t_{v_i}(d_j) - t_{ED}(r_j) - \tau(o_j, d_j) \right) \cdot z_{r_j}^{v_i} \qquad (3)$$

Note that, the arrival time of driver $v_i$ at the pickup location $o_j$ of passenger $r_j$ indicates the actual pickup time. In eq. (3) the waiting time for pickup of passengers is included, and the excess travel time for passenger $r_j$ is calculated as the difference between actual ride-sharing travel time and the fastest travel time.

Our mathematical formulation is extended from the original DARP and is shown below. We assume a public ride-sharing system, which has a societal objective to provide a convenient ride-sharing service and reduce the total vehicle kilometer traveled (VKT) as defined in eq. (4):

$$\min z = \sum_{v_i \in V} \sum_{(a,b) \in A} l_{(a,b)} \cdot x_{(a,b)}^{v_i} + \sum_{r_j \in R} l_{(o_j, d_j)} \left( 1 - \sum_{v_i \in V} \sum_{(a,d_j) \in A} x_{(a,d_j)}^{v_i} \right) \qquad (4)$$

where the first term represents the VKT of all ride-sharing drivers, no matter they are matched or not. The second term captures the VKT of unmatched passengers, which also serves as a penalty for changing to drive-alone of the correspondent unmatched passengers.

The set of constraints for the ride-sharing matching problem is defined in eqs. (5) – (15). Eqs. (5) and (6) are two non-linear definitional constraints. We define the arrival time of driver $v_i$ at each node $b$ as eq. (5):

$$t_{v_i}(b) = \sum_{a:(a,b) \in A} x_{(a,b)}^{v_i} \cdot \left( tt_{(a,b)} + t_{v_i}(a) \right), \forall b \in N \qquad (5)$$

Note that, our arrival time constraint is an equality constraint, which is different from classical DARP formulations. The inequality constraint considered by DARP problems potentially can help to reach better objective values by holding drivers and on-board passengers at current location to fulfill the time constraints at the next stop. However, this not only increases the driver's inconvenience, but also increases actual travel time of on-board peer passengers, and as a result reducing the peer-to-peer ride-sharing service attractiveness. Therefore, we consider the case that after drivers picking up passengers, they depart immediately to their next stops, i.e. the arrival time at node $b$ is equal to the sum of arc travel time $tt_{(a,b)}$ and the arrival time at node $a$.

Constraint (6) defines how the occupancy $Q_{v_i}$ changes at each node the driver visits:

$$Q_{v_i}(b) = \sum_{a:(a,b) \in A} x_{(a,b)}^{v_i} \cdot \left( q_b + Q_{v_i}(a) \right), \forall b \in N \qquad (6)$$

For example, $Q_{v_i}(b)$ indicates that node $b$ is assigned to driver $v_i$ and the occupancy at node $b$ is equal to the sum of the number of passengers pickup or drop-off at node $b$, $q_b$, and the occupancy at node $a$, $Q_{v_i}(a)$.

Constraints (7) and (8) below are conservation constraints. Constraint (7) indicates that the number of picked up passengers for a driver is equal to drop-off passengers for a request $r_j$ at his/her origin and destination.

$$q_{o_j} + q_{d_j} = 0, \forall r_j \in R \qquad (7)$$

Constraint (8) is the flow conservation: it ensures that a driver arriving at a node in the network will also leave the node at the same time period. Moreover, it ensures that if the passenger is picked up, the driver will also have to drop him off at his destination.

$$\sum_{a:(a,b) \in A} x_{(a,b)}^{v_i} = \sum_{c:(b,c) \in A} x_{(b,c)}^{v_i} = z_{r_b}^{v_i}, \forall b \in N_R, \forall v_i \in V \qquad (8)$$

Eq. (9) indicates that each node in the passenger-driver network is visited at most once, i.e. a pickup or drop-off is performed at most once for each passenger:

Yao, Bekhor$$\sum_{v_i \in V} \sum_{a:(a,b) \in A} x^{v_i}_{(a,b)} \leq 1, \forall b \in N_R \quad (9)$$

Eq. (10) ensures that every driver will leave his/her origin and arrive at his/her destination:

$$\sum_{b:(o_i,b) \in A} x^{v_i}_{(o_i,b)} = \sum_{a:(a,d_i) \in A} x^{v_i}_{(a,d_i)} = 1, \forall v_i \in V \quad (10)$$

Constraint (11) is the precedence constraint, which guarantees that each destination node will only be visited after its paired origin is visited:

$$t_{v_i}(d_k) \geq t_{v_i}(o_k), \forall v_i \in V, \text{where } k \in \{v_i, r_j | r_j \in R, z^{v_i}_{r_j} = 1\} \quad (11)$$

Eq. (12) assures that if passengers are assigned to a driver, each pickup node can only be visited after the earliest departure time.

$$t_{v_i}(o_k) \geq t_{ED}(k), \forall v_i \in V, \text{where } k \in \{v_i, r_j | r_j \in R, z^{v_i}_{r_j} = 1\} \quad (12)$$

Together with constraints eqs. (5) and (12), we consider only assigning ride-sharing drivers to passengers who are ready for pickup, i.e. the ride-sharing drivers and on-board peer passengers do not need to wait for picking up the new passenger. For simplicity, we also assume drivers depart from their origins at their desired earliest departure times, $t_{v_i}(o_i) = t_{ED}(v_i)$.

Constraints (13)-(14) indicate level-of-service constraints. The maximum waiting time is set in eq. (13), and the maximum excess travel time is captured by eq. (14).

$$\omega_{r_j} \leq \Omega_{r_j}, \forall r_j \in R \quad (13)$$

$$\delta_k \leq \Delta_k, \forall k \in (V \cup R) \quad (14)$$

The last constraint, eq. (15) ensures at any node on driver's route, the vehicle capacity is never violated, where we also assume all vehicles are initialized with no ride-sharing passengers, $Q_{v_i}(o_i) = 0$:

$$\max(0, q_a) \leq Q_{v_i}(a) \leq \min(c_i, c_i + q_a), \forall v_i \in V, \text{where } a \in \{o_i, d_i, o_j, d_j | r_j \in R, z^{v_i}_{r_j} = 1\} \quad (15)$$

Constraint eq. (15) states that, the minimum occupancy of vehicle $v_i$ is $q_a$ if node $a$ is a pickup node, and 0 otherwise. The maximum occupancy is $c_i + q_a$ if node $a$ is a drop-off node, and equals to $c_i$ otherwise.

**Constraint Linearization**

The arrival time definitional constraint (eq. (5)) and occupancy update constraint (eq. (6)) are non-linear. To allow comparing our approach with mixed-integer program exact solution algorithm, we present the linearization of these two constraints in the following paragraphs.

*Arrival time constraint*

The arrival time constraint (eq. (5)) is linearized by introducing two constants $M^{v_i}_{1,(a,b)}$ and $M^{v_i}_{2,(a,b)}$ for each driver at each link $(a,b) \in A$, and re-written into two inequality constraints eqs. (16) – (17):

$$t_{v_i}(b) \geq tt_{(a,b)} + t_{v_i}(a) - M^{v_i}_{1,(a,b)} \cdot \left(1 - x^{v_i}_{(a,b)}\right), \forall (a,b) \in A, v_i \in V \quad (16)$$

$$t_{v_i}(b) \leq tt_{(a,b)} + t_{v_i}(a) + M^{v_i}_{2,(a,b)} \cdot \left(1 - x^{v_i}_{(a,b)}\right), \forall (a,b) \in A, v_i \in V \quad (17)$$

To simplify the definition expressions, we first introduce the earliest and latest arrival time for both pickup and drop-off nodes as $ET(b)$ and $LT(b)$ respectively:

$$ET(b) = \begin{cases} t_{ED}(b), \forall b \in O \\ t_{ED}(b) + \tau(o_b, d_b), \forall b \in D \end{cases} \quad (18)$$

$$LT(b) = \begin{cases} t_{ED}(b) + \Omega_b, \forall b \in O \\ t_{ED}(b) + \Omega_b + \tau(o_b, d_b) + \Delta_b, \forall b \in D \end{cases} \quad (19)$$

And the validity of constraint (16) and (17) is then ensured by setting $M_{1,(a,b)}^{v_i}$ and $M_{2,(a,b)}^{v_i}$ as following:

$$M_{1,(a,b)}^{v_i} \geq tt_{(a,b)} + LT(a) - ET(b), \forall (a,b) \in A, v_i \in V \quad (20)$$

$$M_{2,(a,b)}^{v_i} \geq tt_{(a,b)} + LT(b) - ET(a), \forall (a,b) \in A, v_i \in V \quad (21)$$

*Occupancy update constraint*

Similarly, the occupancy update constraint eq. (6) is linearized with constant $W_{(a,b)}^{v_i}$ for each driver at each link $(a,b) \in A$, and transformed to:

$$Q_{v_i}(b) \geq q_b + Q_{v_i}(a) - W_{(a,b)}^{v_i} \cdot \left(1 - x_{(a,b)}^{v_i}\right), \forall (a,b) \in A, v_i \in V \quad (22)$$

where $W_{(a,b)}^{v_i}$ is set as following to ensure the validity of constraint (22):

$$W_{(a,b)}^{v_i} \geq \min(c_i, c_i + q_a), \forall (a,b) \in A, v_i \in V \quad (23)$$

**Solution Properties of the Matching Problem**

Madsen et al. (1995) discussed several solution properties of an advanced form of DARP, in which time windows, multiple vehicle capacities, and multiple objectives were introduced. Because of the similarity between our proposed on-demand peer-to-peer ride-sharing matching problem and DARP, our problem is also NP-hard (See Appendix 1 for proof). Thus, these solution properties are reexamined in the context of ride-sharing matching.

**Property 1.** *Capacity Property***:** *If a passenger request can be inserted into a driver's schedule, then the capacity required by the passenger request can be obtained at any of the nodes between its pickup node and the drop-off node.*

Property 1 states that for a feasible passenger request, if there exists a node where the request exceeds the capacity, both the pickup node and the drop-off node are inserted either before or after this node. This property is also valid in the context of ride-sharing matching as defined in eq. (25), in which the occupancy at each node should not violate the capacity constraints.

**Property 2.** *Maximum Travel Time Property***:** *If a passenger request is inserted into the driver's schedule and the maximum travel time is exceeded for the passenger request, then either the pickup node has to be placed later in the schedule, or the drop-off node has to be placed earlier in the schedule.*

We need to extend property 2 in the context of on-demand peer-to-peer ride-sharing matching. In ride-sharing services, we consider drivers as clients and therefore introduce maximum excess travel time constraints also for drivers. The additional time constraints can be seen as adding a fixed node (the driver's destination) to the schedule, where all the passenger requests should be inserted before it. By assuming that drivers leave their origins at their earliest departure time, the maximum travel time for drivers can be defined as the sum of their earliest departure time and their shortest path travel times. And we define maximum excess travel time for passengers to allow a more flexible ride-sharing system.

**Property 3.** *Time Windows Property***:** *If the upper bound on a time window of pickup node or drop-off node is violated, then the time window will still be violated if the pickup node or drop-off node is moved later in the driver's schedule.*

In our formulation, all of the time constraints are actually time window constraints, in which the lower bounds are the earliest departure times. The upper bound of a pickup node is defined by the sum of its earliest departure time and the maximum waiting time. The upper bound of a drop-off node is equal to the sum of its earliest departure time, shortest path travel time, and maximum excess travel time. Property 3 states that, if the upper bound is violated at certain location in the driver's schedule, we can skip inserting the node into locations later than that, thus reducing the search space and improving the efficiency.

**SOLUTION APPROACH**

The on-demand peer-to-peer ride-sharing problem defined in the previous section is very complex to solve in large networks. Therefore, there is a need to reduce the problem complexity. The proposed solution scheme decouples the optimization problem into simpler subproblems, which allows handling of larger problem instances. We propose to divide the problem into two main levels, according to the following framework presented in Figure 2.

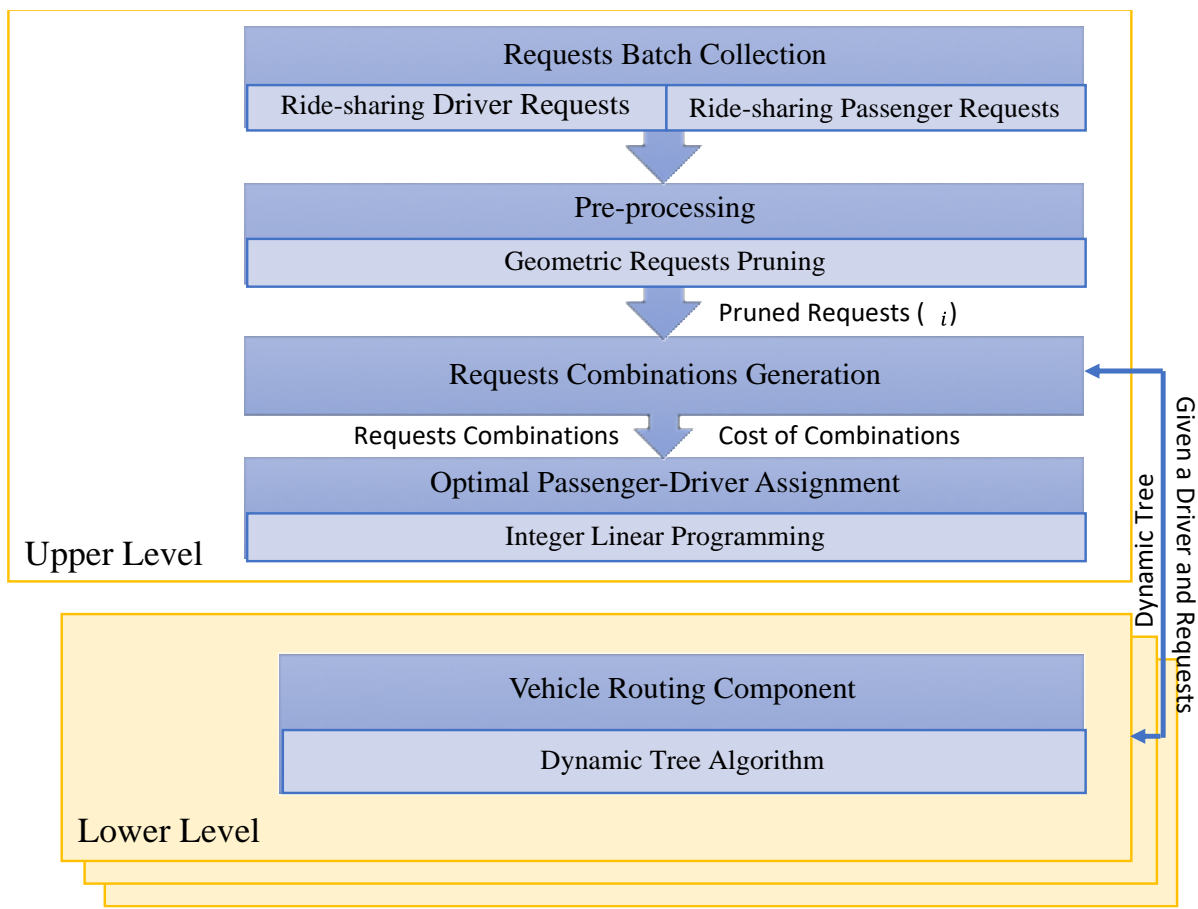

Figure 2 Solution framework for the on-demand peer-to-peer ride-sharing problem

Upper Level:

- Request Batch Collection: passenger and driver requests are collected in a batch for a predefined period of time (e.g. 5 min) and passed to the next step

- Pre-processing of drivers' and passengers' requests: we prune the requests for passengers and drivers that fall beyond the maximum waiting time and/or maximum excess travel time constraints.

- Requests Combinations Generation: constructing feasible passenger combinations for all the drivers by solving the lower level problem. This allows us to compute the costs (objective values) of the combinations.

- Solving the optimal passenger-driver assignment problem via integer linear programming.

Lower Level:

- For a single driver, solve the vehicle routing problem (VRP) using dynamic tree (explained later): find the optimal route for given passenger requests with respect to the objective functions and return the objective values to the upper-level problem. If a feasible route does not exist, return "Infeasible".

Our solution approach is driver-oriented, in which the Pre-processing step extracts the candidate requests for each driver $v_i$. The above framework allows to compute independently the Pre-processing step and Requests Combinations Generation step for each driver, which can be implemented in a parallel fashion.

The ride-sharing matching solution framework contains mainly four steps. First, the passenger and driver requests are grouped respectively for a predefined period of time. Then the Pre-processing step efficiently extracts the candidate passenger requests. Next, the Request Combinations Generation step applies Dynamic Tree Algorithm to combine the feasible passenger requests and to find the driver's service sequences. The last step corresponds to the solution of the optimal passenger-driver assignment problem using integer linear program (ILP). The following subsections describes in detail the main algorithm steps.

**Pre-processing – Geometric Requests Pruning**

This step selects passenger requests that jointly satisfy drivers' and passengers' spatial proximity constraints, and assign these candidate passenger requests to the corresponding drivers. The method proposed by Masoud and Jayakrishnan (2017) for the single-passenger multi-hop ride-sharing problem is further extended to accommodate the multi-passenger multi-driver problem.

The main idea of the pre-processing step is to use maximum waiting time constraints and maximum driving time constraints to select the candidate passengers-driver pairs that may satisfy these constraints, in which both time constraints limit the accessibility of the driver with respect to picking up passengers.

We define the accessible region of the driver as an ellipse, where the focal points are the origin-destination locations of the driver, and the transverse diameter is a distance upper bound with respect to the driver's maximum excess travel time. Only if both the pickup and drop-off locations of passenger requests are within the accessible region, these requests are potentially feasible to the driver. If either the request's pickup or drop-off location is outside the accessible region, assigning these requests will violate the driver's maximum excess travel time constraint.

Note that these potential requests may not be feasible, even both their pickup and drop-off locations are within the accessible region. This is because the actual travel times on the road network are typically higher than the travel times calculated by Euclidean distance. Moreover, if additional pickups/drop-offs are executed, the resulting driver's service schedule may also exceed the maximum excess travel time constraints. The feasibility of assigning these candidate requests to drivers is validated by solving a lower level VRP, which will be presented later. We define these potential requests for driver $v_i \in V$ as $R'_{v_i}$.

To further reduce the solution search space, we define the reachable pickup region of the passenger as a circle where the pickup location is the center of the circle and calculate the radius by assuming a maximum speed and maximum waiting time. The reachable pickup region selects the drivers who are able to pick up this passenger request while satisfying the maximum waiting time constraint. We define these potential drivers for passenger $r_j \in R$ as $V'_{r_j}$.

The resulting candidate passenger set $R_{v_i}$ of driver $v_i$ should jointly satisfy both the maximum excess travel time constraints for the driver, and the maximum waiting time constraints for the passenger requests, i.e. the reachable pickup region includes the driver's current location and the driver's accessible region includes the request's pickup location (Figure 3 (c)):

$$R_{v_i} = \left\{r_j \mid \forall r_j \in R'_{v_i}, v_i \in V'_{r_j}\right\} \tag{24}$$

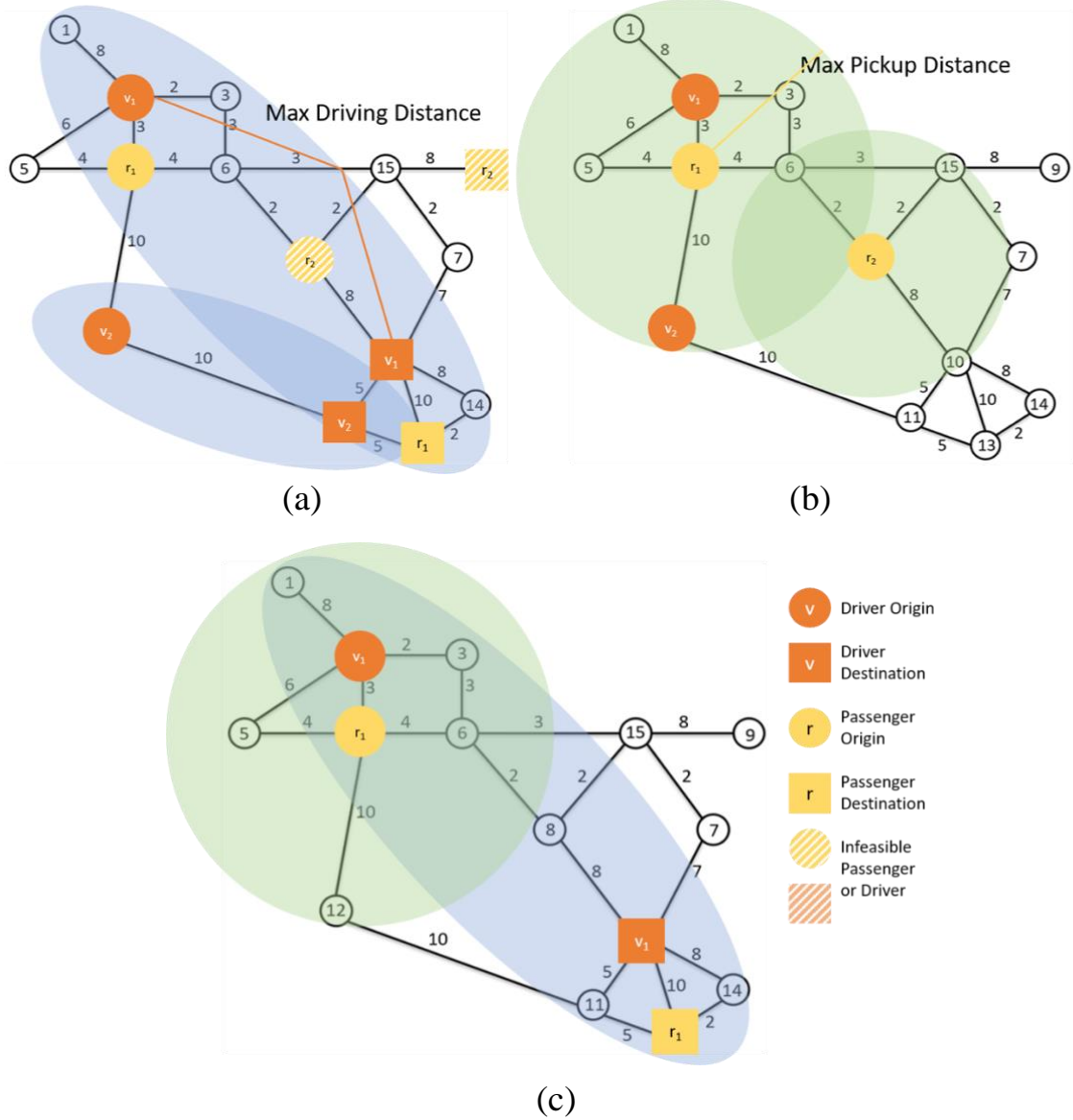

Figure 3 Pruning example

Figure 3 shows an example of 2 drivers $(v_1, v_2)$ and 2 passenger requests $(r_1, r_2)$ in the Mandl's network. The pruning for drivers is shown in Figure 3(a), in which the blue ellipses represent the accessible region of the drivers. We can see that for driver $v_2$ neither of the requests is within the accessible region, while for driver $v_1$, request $r_1$ is in his/her accessible region. Note that the destination of request $r_2$ falls outside the accessible region of driver $v_1$. Therefore, the drivers' potential request set is $R'_{v_1} = \{r_1\}$ and $R'_{v_2} = \emptyset$.

The pruning for passengers is shown in shown Figure 3(b). In this example, passenger $r_1$ has a larger reachable pickup region, and both drivers $v_1, v_2$ are within the region. As for passenger $r_2$, neither driver is within the reachable pickup region. In this case, the system will reject this request. Therefore, the passengers' potential driver set is $V'_{r_1} = \{v_1, v_2\}$ and $V'_{r_2} = \emptyset$.

Finally, the candidate requests set for each driver are shown in Figure 3(c). Since passenger request $r_1$ and driver $v_1$ intersects with each other in their potential set and there is no intersection with $r_2$ and $v_2$, the candidate request set will be $R_{v_1} = \{r_1\}$ and $R_{v_2} = \emptyset$. These candidate requests are passed to the Request Combinations Generation step, described in the next subsection. The feasibility of the request combinations is checked by finding a route that satisfies both driver's and passengers' constraints. Since many combinations are needed to be checked, an efficient VRP algorithm should be developed, we present this algorithm after the request combination generation subsection.

**Request Combinations Generation**

Given a candidate request set $R_{v_i} = \{r_1, \dots, r_n\}$ with $n$ requests, we need to determine which requests can be served together by driver $v_i$ with respect to the constraints, by checking different request combinations $p_{v_i}$.

We implement an efficient request combination generation procedure similar to Alonso-Mora et al. (2017), by translating spatiotemporal sharing problems into a graph-theoretic framework (Santi et al., 2014). The procedure first checks pairwise shareability, that is, if driver $v_i$ can serve each one of the requests $r_j \in R_{v_i}$ alone and satisfying all the time constraints.

Next, it finds the feasible cliques (a subset of nodes of a graph such that every two distinct nodes in the clique are adjacent) incrementally in size. For all requests that can be served by driver $v_i$ alone, we check if the combinations of 2 requests can be served by this driver with respect to the constraints. We further check combinations of size 3, 4, …, $k$ by joining feasible combinations with size $(k - 1)$. Consequently, different requests and request combinations are considered and verified. The essence of this approach is based on the following Lemma for computing the feasible combinations:

"A request combination $g_{v_i}$ can be feasible only if there exists a driver $v_i$ for which, all the requests $r \in g_{v_i}$, the sub-combinations $g'_{v_i} = g_{v_i} \backslash r$ are feasible" (adapted from Alonso-Mora et al., 2017).

The feasibility of a combination $g_{v_i}$ is verified through solving the lower-level problem (VRP) with respect to the constraints. If a feasible optimal route is found, then the cost (objective value) of the combination is also calculated. Otherwise, the combination is infeasible.

Instead of finding routes from scratch, we implement the dynamic tree algorithm for solving the VRP that makes use of the previously computed routing solutions. According to the Lemma, all the sub-combinations $g'_{v_i} = g_{v_i} \backslash r$ are feasible. In order to find a convenient route for the driver with respect to the new combination $g_{v_i}$, we will only need to add any request $r \in g_{v_i}$ to the respective dynamic tree of the sub-combinations $g'_{v_i}$ via dynamic tree algorithm. In the following subsection, we describe the Dynamic Tree Algorithm step in detail.

**Dynamic Tree Algorithm**

We propose a dynamic tree algorithm to solve the ride-sharing VRP. The algorithm augments the vehicle schedule based on previously calculated schedules. The fundamental idea of this algorithm is to keep track of the feasible VRP solutions for a given set of passenger requests, using a dynamic tree structure. Remind that infeasible request combinations will remain infeasible when considering service sequences with new requests. Therefore, given the driver's current location, his/her destination, and a set of temporally assigned requests, we only need to insert the new requests into the existing dynamic tree to obtain new feasible solutions.

We will first introduce the characteristics of dynamic trees. The dynamic tree maintains the information about driver's current location, destination, pickup and drop-off sequences of the previously assigned requests, driver's scheduled arrival time at each node in the sequence, and the time constraints. The dynamic tree handles the insertion of new passenger requests into the current sequence and obtains the optimal route. The dynamic tree also updates the driver's schedule with respect to the driver's current location. When the driver reaches a node in the sequence, the subsequence of the schedule before this node becomes obsolete, and thus the subsequence is removed from the dynamic tree. These characteristics justify the suitability of dynamic trees to handle on-demand peer-to-peer ride-sharing matching problem.

For a given dynamic tree $T$, we define the current location (origin) of driver as root $l$ of $T$, the destination of driver as $e$. The level of a node is defined as the number of edges between the node and the root, denote $T[n]$ as the nodes of tree $T$ at level $n$. Note that the level of a node also indicates its spatiotemporal proximity to the root. We also define a child as a node $s$ directly connected to another node $k$ when moving away from the root, denote node $k$ as the parent of node $s$, and a group of nodes with the same parent as the siblings. A subtree of node $s$, $T(s)$, is defined as a tree which is a child of $s$ and is a portion of $T$ under node $s$. And the new tree generated by the algorithm is defined as $T'$.

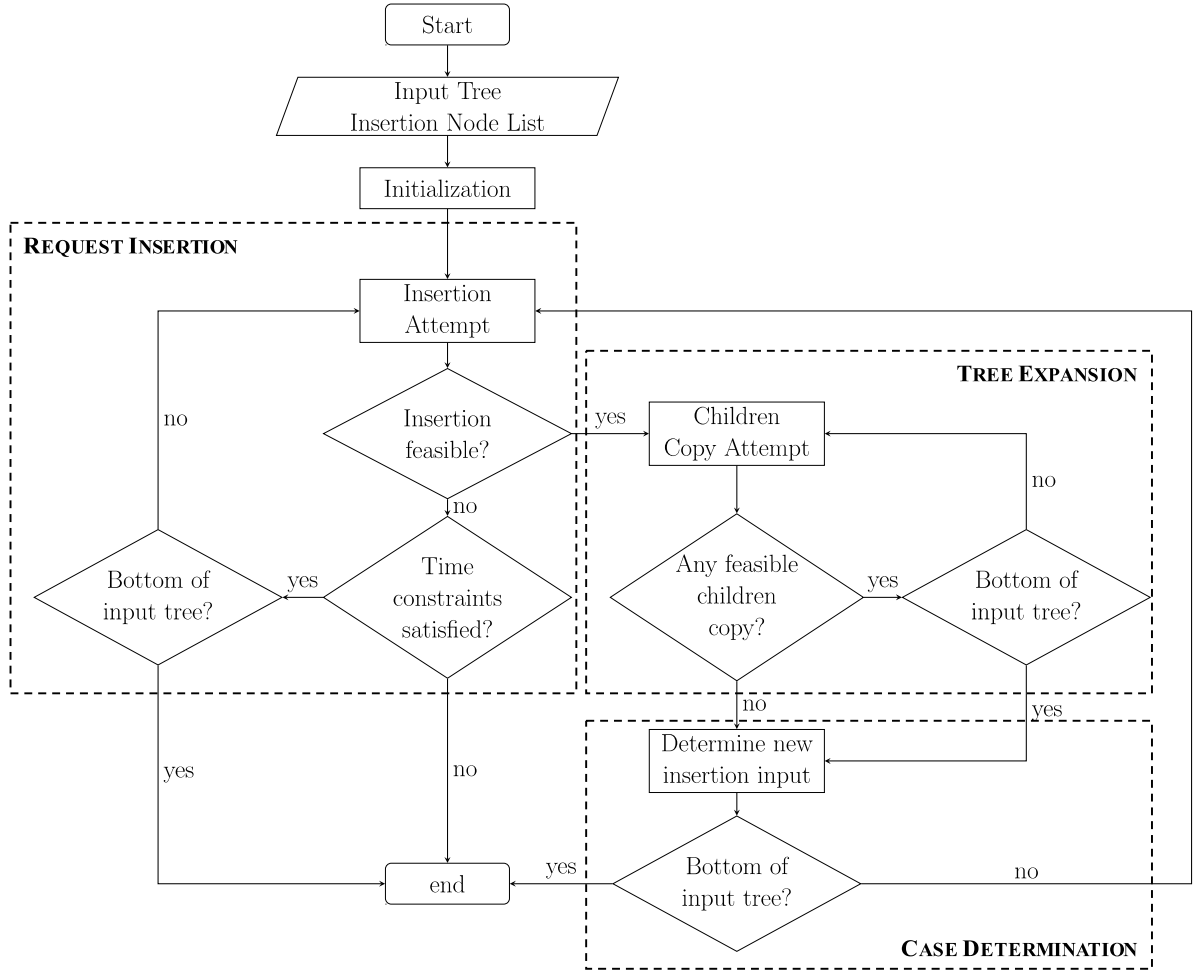

Figure 4 Dynamic tree algorithm flowchart

Figure 4 outlines the proposed dynamic tree algorithm. The algorithm requires input of a driver's schedule (the initial dynamic tree $T$), the insertion node list $N = (o, d)$ where $o$ and $d$ are respectively the pickup and drop-off nodes. The nodes in $N$ will be inserted in order, i.e. we will first insert the origin and construct a new tree that includes the origin, after which the destination will be inserted under the origin in the new tree. This insertion order satisfies the precedence constraint. Furthermore, the insertion starts under the root of $T$ and moves downwards in $T$. According to the solution properties, we efficiently determine the feasibility of the node insertion.

After initialization, the algorithm starts inserting nodes by calling REQUEST INSERTION procedure. If the insertion is feasible, TREE EXPANSION procedure is called to construct a new tree that includes the inserted node. After the execution of the TREE EXPANSION procedure, CASE DETERMINATION procedure decides the continuation of the algorithm. We now describe these three main procedures in detail.

*Request Insertion Procedure*

The REQUEST INSERTION procedure essentially needs to determine which edges of the given tree $T$ can accommodate the insertion of pickup and drop-off nodes, respectively. A feasible insertion should satisfy both capacity and time constraints. In case the insertion is feasible, TREE EXPANSION procedure constructs a new tree that includes the inserted node and satisfies all the constraints. Otherwise, the algorithm needs to find an alternative insertion downwards in the tree. However, this searching procedure is very time consuming. According to the solution properties (explained below), we can determine if it is feasible to continue the searching, which saves extra computational time. As a result, the proposed dynamic tree algorithm is very efficient.

If the insertion, for either pickup or drop-off node, is infeasible due to violation of time constraints, we can stop the procedure. According to Solution Property 3, if this insertion is infeasible, the time window constraints remain violated downwards in the tree. However, if only the capacity constraint is violated and the inserted node is a pickup node, it is possible to insert it later in the schedule. For instance, if the capacity constraint for the pickup node is violated under current parent $p$, it still may be feasible to insert this node after some drop-offs occur. In this case, we keep the insertion node list $N$, and recursively insert the pickup node downwards in the tree. Note that if the capacity constraint for the drop-off node is violated, we can also stop the insertion procedure due to Solution Property 1. The solution properties 1 and 3 can significantly reduce the number of insertions in this procedure with respect to the given constraints and improve the performance of the algorithm. The proposed procedure above, combines Alonso-Mora et al. (2017) general framework with the dynamic tree approach of Huang et al. (2017).

*Tree Expansion Procedure*

Given a parent $p$ of the inserted node and an input tree $T$, TREE EXPANSION procedure needs to verify the feasibility of the nodes in the subtree $T(p)$ after the insertion, and copies the feasible nodes to a new tree in a recursive manner. The procedure starts from the children of $p$ and moves downwards, in which subtrees of feasible children will be served as input to the next copy execution. A new tree $T'$ generated by TREE EXPANSION procedure is feasible if it has at least one leaf node as the driver's destination, and all constraints are satisfied. The procedure terminates when there is no feasible children copy, or it reaches the bottom of the input tree $T$.

Note that in TREE EXPANSION procedure, it is unnecessary to check the capacity constraint. The capacity constraint violations can only occur when a pickup node is inserted. According to Solution Property 1, we will only need to insert the drop-off node before this violating node. Moreover, we also employ Solution Property 3 to help us determine the continuation of the copy execution. If the time constraints are violated, the procedure can skip the subtree of current node and move to its siblings.

*Case Determination Procedure*

After the execution of TREE EXPANSION procedure, there are four possible cases as follows:

1. If the inserted node is a pickup node and the new tree $T'$ is feasible, it means the new pickup sequence with the inserted pickup satisfies all the constraints. The corresponding drop-off node is inserted under the pickup node in the new tree $T'$ to find a feasible driver schedule. The new tree $T'$ is served as an input tree $T$ to the REQUEST INSERTION procedure for the drop-off node insertion.

2. If the inserted node is a pickup node but the new tree $T'$ is infeasible, the pickup node is inserted under the child or sibling of the current parent (i.e. move downwards in input tree $T$). The algorithm terminates if it reaches the bottom of $T$.

3. If the inserted node is a drop-off node and the new tree $T'$ is feasible, there exists at least one feasible driver schedule for the given passenger requests. The REQUEST INSERTION procedure continues insertion for both pickup and drop-off nodes, the insertion occurs under the child or sibling of the current parent of the pickup node.

4. If the inserted node is a drop-off node but the new tree $T'$ is infeasible, i.e. we do not find a feasible driver schedule with respect to this pickup sequence, the pickup node insertion is failed. The new tree $T'$ is discarded, and both pickup and drop-off nodes are inserted by the REQUEST INSERTION procedure if there remain other insertion points downwards in tree $T$. Otherwise, the algorithm terminates.

*Example*

We illustrate the dynamic tree algorithm depicted in Figure 5 as follows: consider a case that a driver $v_1$ is willing to take 2 passengers, in which passenger $r_1$ has already been assigned but not yet picked up, and a new passenger request $r_2$ arrived in the system. The corresponding initial input tree $T$ is shown in Figure 5(a), in which there are 3 possible insertion points for origin node for $r_2$.

First, REQUEST INSERTION procedure tries to insert the origin node for $r_2$ at insertion point 1. Since the driver is willing to take 2 passengers, and we only have 2 requests, it is unnecessary to check for capacity constraints in this case. Suppose the earliest pickup time and maximum waiting time constraint of origin node for $r_2$ at insertion point 1 is satisfied, the REQUEST INSERTION procedure generates a new tree $T'$ as shown in Figure 5(b).

Next, the TREE EXPANSION is executed after the origin node for $r_2$ insertion, as shown in Figure 5(c). The TREE EXPANSION procedure copies the subtree of the root in the input tree $T$ (dashed line box in Figure 5(a)) under the origin node for $r_2$ in $T'$.

Assume that the additional waiting time endured by $r_1$ is still within his/her maximum waiting time constraint, it is feasible to copy the origin node for $r_1$ to $T'$. However, due to the extra waiting time, the maximum excess travel time constraint for $r_1$ is violated, it is infeasible to copy the destination node for $r_1$ to $T'$. According to the Solution Properties, the TREE EXPANSION procedure is terminated due to violation of time constraints (indicated by the crossing lines in Figure 5(c)). Note that the driver's destination $d_v$ is not copied to $T'$ either, thus the resulting new tree $T'$ is not feasible. We need to discard $T'$ and try to insert the origin node for $r_2$ at point 2.

Now, suppose the REQUEST INSERTION of origin node for $r_2$ at insertion point 2 is feasible and there is a feasible new tree after the TREE EXPANSION procedure. The resulting new tree $T'$ is shown in Figure 5(d). Since we found a feasible pickup node insertion, we turn now to find a feasible insertion for the drop-off node.

The destination node for $r_2$ is inserted under the origin node for $r_2$ in the new tree $T'$, in which there are 2 potential insertion points (shown in Figure 5(d)). The destination node for $r_2$ is inserted firstly at point 4. Assume the maximum excess travel time constraint is satisfied for destination node for $r_2$ at point 4. The algorithm attempts to copy the remaining nodes in the tree, in this case only the driver's destination $d_v$, into $T'$. The same process is repeated for insertion point 5, and assume all the constraints are satisfied, a feasible new tree $T'$ with respect to insertion of origin node for $r_2$ at point 2 is obtained as shown in Figure 5(e).

Lastly, suppose the insertion of origin node for $r_2$ is also feasible at insertion point 3, and all the constraints are satisfied for the destination node for $r_2$ and driver's destination $d_v$ under pickup node $o_2$ at insertion point 3. The algorithm outputs the final new tree $T'$ that includes all feasible driver schedules with the new request $r_2$ (Figure 5(f)).

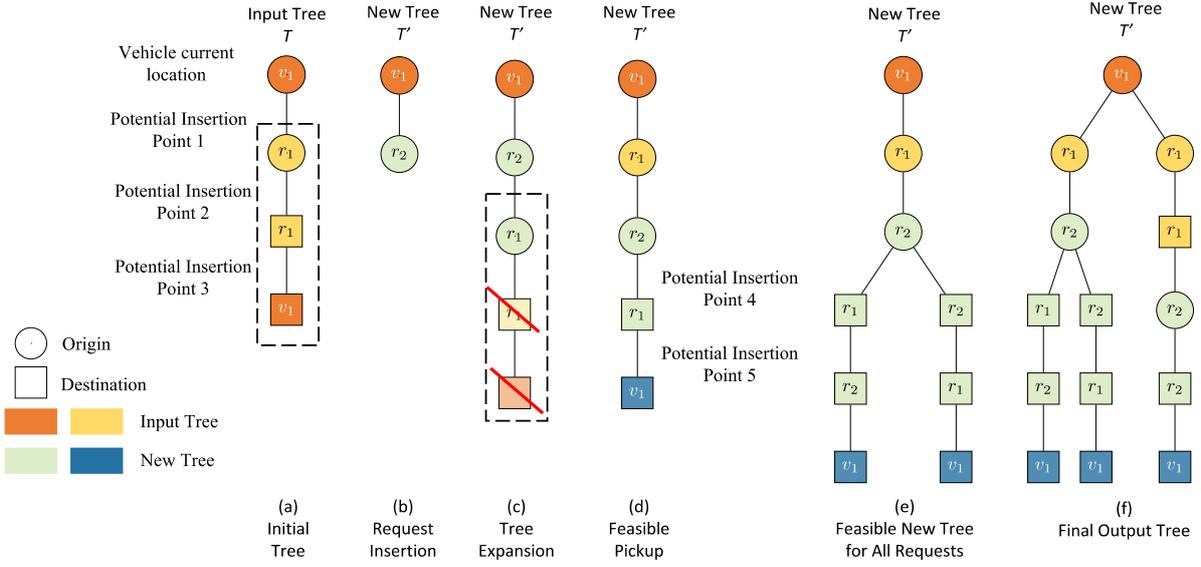

Figure 5 Example of dynamic tree algorithm steps

Note that the insertion sequence of different requests does not influence the objective function values. For a given set of requests, we attempt to insert each request to all feasible locations in the dynamic tree. For example, Figure 5(b) and (c) shows that the dynamic tree algorithm considers the case that $2^{nd}$ inserted request $r_2$ being picked up before picking up the $1^{st}$ inserted request $r_1$, and the cost (objective function value) of these two requests is obtained from all the feasible insertion locations (Figure 5(f)).

The dynamic tree algorithm provides the feasible solutions and objective function values of all combinations generated by the Request Combination Generation step. The values are then used in the following step that assigns passenger combinations to drivers.

**Optimal Passenger-Driver Assignment**

In this step, we formulate the optimal passenger-driver assignment problem as an integer linear programming problem (ILP). The Request Combination Generation step provides the feasible request combinations between drivers and passengers with their objective function values obtained from the dynamic tree algorithm. The simplified ILP formulation needs additional definitions. We denote the total number of request combinations as $n$, the decision vector X (of size n), in which each element $x(i) \in \{0,1\}$. If a combination is assigned then $x(i) = 1$.

We define the objective value vector of passenger request combinations as $\Gamma$ of size n, in which each element represents the objective value of the optimal driver schedule for this combination of passenger requests. The request-combination incident matrix is defined as $\Phi$ of size $[|R|, n]$, in which element $\varphi(i, j)$ is equal to 1 if the request $i$ is included in passenger request combination $j$. The driver-combination incident matrix is defined as $\Psi$ of size $[|V|, n]$. If driver $i$ is associated with passenger request combination $j$, then $\psi(i, j) = 1$, otherwise $\psi(i, j) = 0$.

The above definitions simplify our formulation in eqs. (1) – (23) to the following ILP mathematical formulation:

$$\min z = \Gamma \cdot X \qquad (25)$$

subject to:

$$\sum_{j=1}^{n} \varphi(i, j) \cdot x(j) \leq 1, \forall i \in [1, |R|] \qquad (26)$$

Yao, Bekhor$$\sum_{j=1}^{n} \psi(i,j) \cdot x(j) \leq 1, \forall i \in [1, |V|] \tag{27}$$

$$x(j) \in \{0,1\}, \forall j \in [1, n] \tag{28}$$

Eqs. (25) – (28) are equivalent to the original formulation eqs. (1) – (23). In this formulation, capacity, precedence, and time constraints are purposely omitted, since these constraints are satisfied by the lower-level vehicle routing problem. Similar to eqs. (9) and (10), eqs. (26) and (27) impose that for any passenger request or driver, at most one combination is assigned.

The above formulation is very similar to the maximum weighted bipartite matching problem and can be solved efficiently by existing ILP solvers.

**Discussion of Complexities**

- Pre-processing – Geometric Requests Pruning

    The Pre-Processing step first requires computations of Euclidean distances between drivers and passengers, in which the complexity is scaled linearly with the number of drivers and passengers, and in the worst case the complexity follows $O(|V||R|)$. Next, feasible potential passenger requests should be selected, in which the complexity is of order $O(|V||R|\log(|V||R|))$ in the worst case. We will examine the effectiveness of the Pre-processing and showing the comparison for computation with and without the pre-processing step.

- Dynamic Tree Algorithm

    The dynamic tree algorithm exhaustively inserts the potential requests into the driver's schedule to find the optimal routing solutions. Suppose the maximum number of passengers assigned to the drivers is set to be $m$, in the worst case (i.e. with very loose capacity and time constraints), the algorithm needs to iterate over $\frac{(2m)!}{2^m}$ permutations. And for each permutation, the algorithm needs to verify the feasibility of the insertion of pickup and drop-off nodes with respect to all the constraints, which runs $(2m + 1)$ times.

    The total complexity of the dynamic tree algorithm is $O\left(\frac{(2m)!}{2^m} \cdot (2m + 1)\right) = O\big((2m)!\big)$ in the worst case. However, capacity and time constraints are relatively strict in practice due to the dynamic nature of the proposed ride-sharing system, the dynamic tree algorithm is well suited for solving the problem and run efficiently.

- Request Combinations Generation

    This step finds all the feasible request combinations, in the worst case, the complexity is of order $O(|V||R|^m)$, in which all passenger requests are feasible to all drivers. For each combination, a convenient route is found via dynamic tree algorithm. However, in the practice, the number of feasible combinations is orders of magnitudes lower and it is related to the maximum number of passengers assigned to the drivers .

**NUMERICAL RESULTS**

This section presents two sets of examples. First, we compare objective function values and runtime performances between our proposed algorithms and branch and cut exact solution algorithm using simple grid networks. Next, we examine the efficiency of our proposed on-demand peer-to-peer ride-sharing matching algorithm using the well-known Winnipeg network, and conduct sensitivity analysis to validate the ride-sharing system performance and to identify important factors that influence the algorithm runtimes. We implement the proposed dynamic tree algorithm using Python 3.7 programming language, Gurobi 8.1.1 as the ILP solver, and CPLEX 12.10 as the mixed integer programming (MIP) solver on a 6-core PC with 16 GB RAM.

**Grid Network Examples**

The following examples are generated by randomly choosing the coordinates of pickup and drop-off nodes in a [-10,10] x [-10, 10] grid network according to a uniform distribution. All vehicles depart from the same origin and arrive at the same destination (0, 0). The following constraints are considered in these examples: (a) All vehicles have capacity of 3; b) the maximum travel time for all participants is 240 minutes; c) the maximum waiting time for pickup is 15 minutes; and d) the maximum travel distance is 30 km (assuming average speed of 60km/h). The comparison of objective values and runtime performance between branch and cut exact algorithm and dynamic tree algorithms are presented in Table 1.

Table 1 Objective values and runtime performances comparisons

|  | CPLEX Branch and Cut | | Dynamic Tree | |
| --- | --- | --- | --- | --- |
| **Number of vehicles - Number of passengers** | **Obj. Value** | **Runtime [s]** | **Obj. Value** | **Runtime [s]** |
| 4-10 | 200.89 | 16.30 | 200.89 | 0.67 |
| 4-16 | 290.09 | 222.96 | 290.09 | 5.80 |
| 4-18 | 294.72 | 2719.94 | 294.72 | 14.35 |
| 4-20 | 298.56 | 14046.88 | 298.56 | 45.65 |
| 5-24 | 487.33 | 14227.63 | 487.33 | 61.28 |

Using the linearized formulation (eqs. (1) – (23)), we solve the MIP with branch and cut exact algorithm in Cplex. Results show that, the dynamic tree algorithm reaches the same objective function values with much shorter runtimes in all examples. This is because our dynamic tree algorithm utilizes the DARP solution properties to quickly determine the continuation of the insertions. Another noticeable reason is related to the Request Combination Generation procedure, which is based on graph cliques. This procedure avoids checking request combinations that are not feasible, thus reducing considerably the number of possible matchings.

Results suggest that exact branch-and-cut algorithm is unable to handle larger examples (note that the biggest case includes only 5 drivers and 24 requests). Clearly, the exact algorithm is not suitable for on-demand peer-to-peer ride-sharing matching, in which a large number of requests arrive continuously in the system, which brings additional complexity to the problem. In the following subsection, we show that our proposed dynamic tree algorithm handles a real size network and a relatively large number of requests.

**Winnipeg Network Example**

In this subsection, we conduct a series of numerical tests using the well-known Winnipeg network to examine the efficiency of our proposed on-demand peer-to-peer ride-sharing matching algorithm. The Winnipeg network consists of 154 zones, 1,067 nodes, 2,535 links, and 4,345 origin-destination (OD) pairs. The total hourly demand of the Winnipeg matrix is 54,459 trips (Bekhor et al., 2008).

To obtain insights of the algorithm steps, we assume a default setting as follows:

- 3,000 ride-sharing participant requests, in which 1,000 of them are willing to be drivers (denoted by $V$), and the remaining 2,000 are passenger requests (denoted by $R$).

- The origins and destinations of the requests are randomly sampled from the given OD trip distribution.

- Ride-sharing vehicles have capacity of 4 and are empty at the beginning of the time period.

- The maximum excess travel time constraints for passengers and drivers are assumed as:

$$\Delta_k = 20\% \cdot \tau(o_k, d_k), \forall k \in (V \cup R) \quad (29)$$

- The maximum pickup waiting time is assumed as:

$$\Omega_{r_j} = 50\% \cdot \Delta_{r_j}, \forall r_j \in R \quad (30)$$

- 100 times of random OD draws for each setting

Note that our algorithm can assign any number of passengers to a driver, if there are available seats. In this paper, we assume the maximum size of passenger request combinations equal to 4, which means that the algorithm will assign no more than 4 passengers to a driver in a time period.

**Matching Results**

Figure 6 exemplifies the main algorithm steps in the Winnipeg network. In this example, there are 1,381 different OD pairs for the 2,000 passenger requests. The requests are shown in Figure 6(a), in which pickup nodes are in blue and drop-off nodes are in orange. A specific driver is marked in yellow, and the corresponding time-space ellipse is marked in blue.

The Preprocessing step (geometric pruning procedure) tries to exclude the infeasible requests for this driver. As shown in Figure 6(b), there are only 70 different OD pairs after this step, in which potentially feasible pickup nodes are marked in red and drop-off nodes are marked in green. We can see that the potential pickup points are closer to the driver's origin and most drop-off nodes are closer to driver's destination.

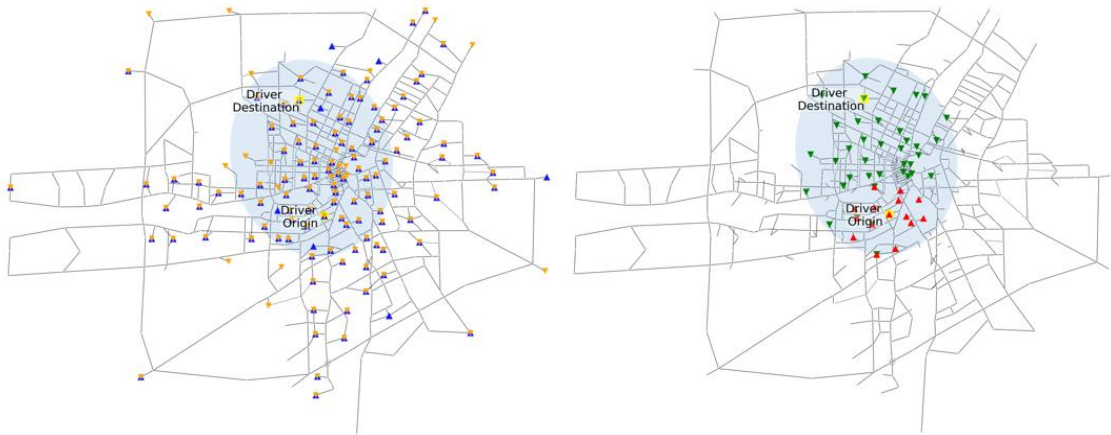

(a) All requests        (b) Potential requests after pruning

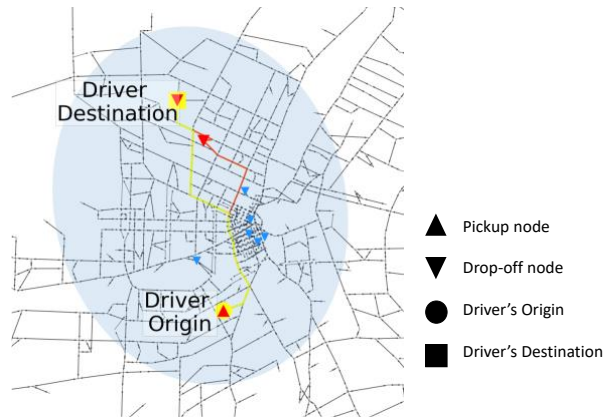

(c) Feasible requests and matching result

Figure 6 Algorithm results

The matching result for this specific driver is shown in Figure 6(c), where assigned requests are marked in red and unassigned feasible requests are in blue. The driver's shortest path is marked in yellow, and the detour path is marked in red. For this specific driver, there are 4 pickups at driver's origin, 1 drop-off at driver's destination, and 3 other drop-offs on the way. We can see that passengers with longer travel distances and closer to the driver's destination are assigned to this driver, which coincides with our objective to maximize the VKT saved. Note that the unassigned passenger requests may be assigned to other drivers and result in larger total VKT saved.

*OD Distribution*

We performed 100 replications, in which participant requests are randomly sampled to assess the impact of OD distribution in the network. We define the match success rate as follows:

$$\text{Match Success Rate} = \frac{N_{MV} + N_{MR}}{|V| + |R|}\% \tag{31}$$

where $N_{MV}$ and $N_{MR}$ are the total number of matched drivers and passengers respectively. The matching rate is relatively consistent with respect to OD distribution. For the constraint sets defined in eqs. (24) and (25), the average match success rate is 63.21% with standard deviation 1.26%.

The total runtime against the total number of feasible request combinations (for each of the 100 replications) is shown in Figure 7. The runtime includes pre-processing, request combinations, and ILP computation time.

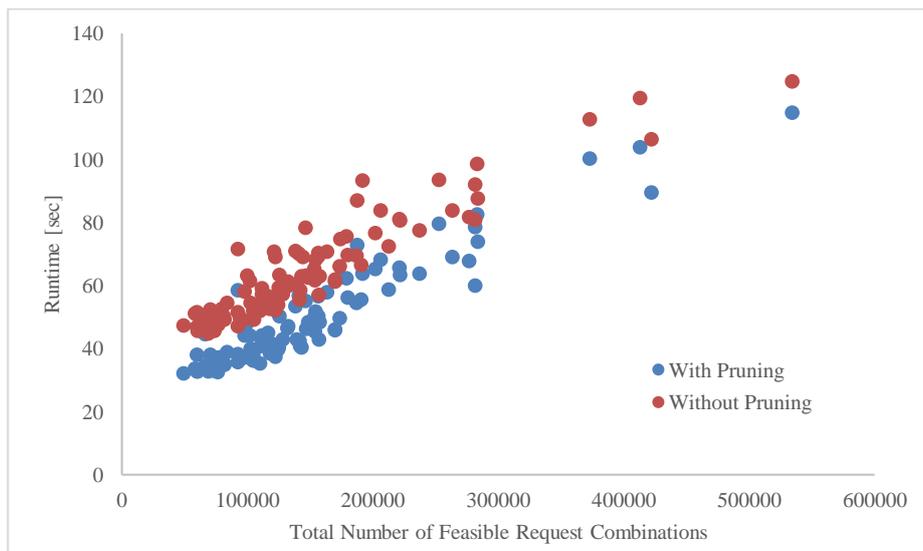

Figure 7 Total runtime for each of the 100 random draws

Each point in Figure 7 represents a random draw, which may have different number of feasible passenger requests combinations. As shown in Figure 7, the runtime is almost linear with respect to the number of feasible passenger requests combinations.

The figure shows that the OD distribution matters to the total runtime, due to the differences in the total number of feasible request combinations. The average total runtime is 49.02 sec with a standard deviation of 16.38 sec. Runtime breakdown by algorithm steps: average Pre-processing time is 0.09 sec, average Request Combinations Generation step is 42.25 sec, and ILP solving time is 6.61 sec.

*Impact of Pruning*

In Figure 7, runtimes of the algorithm with and without pruning are shown. The objective function values, feasible passenger requests combinations, and matching results are identical with or without pruning for all 100 replications, which indicates the pruning process only eliminates the infeasible search space and does not reduce the feasible region.

The average runtime for the algorithm without pruning process is about 63.64 sec with a standard deviation of 16.72 sec. Compare to it, the algorithm with pruning saves about 14.62 sec for each run while the pruning time is only 0.09 sec on average. We also expect that with a larger number of participant requests, the pruning process can help further improve the computational efficiency.

*Ride-sharing Impact on the Network*

To illustrate the impact of the ride-sharing service on the network, we present the VKT savings and VHT savings compared to the base case (no ride-sharing). The results for the default setting are illustrated at the network level in Figure 8, and summarized in Table 2.

The original vehicle trips of matched ride-sharing passenger and driver are combined into ride-sharing vehicle trips, i.e. their original vehicle trips are removed from the origin-destination matrix. While the majority of private vehicle trips are drive-alone trips and are remained in the matrix. Note that unmatched passengers and drivers are reassigned to the network as drive alone. And a traffic assignment is run to obtain the VKT and VHT. We are expecting savings in this specific setting, in which there are 1000 drivers and 2000 passengers with relatively strict excess travel time constraints.

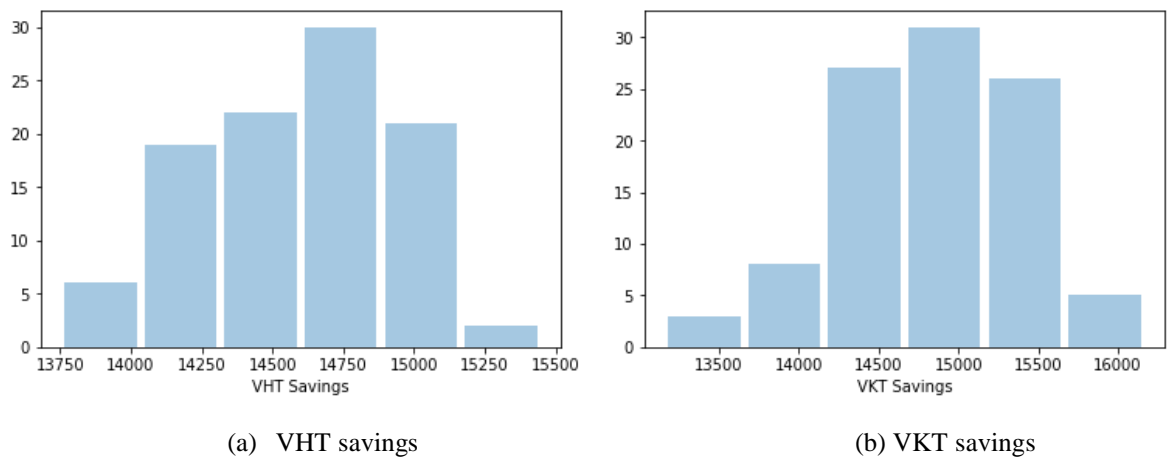

(a) VHT savings  (b) VKT savings

Figure 8 VHT (a) and VKT(b) savings frequency count plot for each of the 100 random draws

Table 2 Summary of ride-sharing impacts on the network

|  | **Base Case** | **Average (100 Replications)** | **Difference** |
|---|---|---|---|
| Vehicle Trips | 54,459 | 53,291 | 2.14% |
| VKT | 720,902 | 706,047 | 2.06% |
| VHT | 706,567 | 692,223 | 2.03% |

Table 2 indicates for the default setting, there are 2.14% trip savings, 2.06% VKT savings, and 2.03% VHT savings compare to the base case (no ride-sharing), with considering only around 5% percent of the vehicle trips for ride-sharing. Notice that, VKT and VHT savings are relatively smaller than the vehicle trip savings, because our proposed ride-sharing system allows drivers to detour for pick up and drop off passengers with different origins and destinations.

**Sensitivity Analysis**

This section performs a sensitivity analysis with respect to selected parameters that affect the algorithm performances, matching statistics, and network impact. For simplicity, we discuss driver supply, time constraints, and capacity constraints. Furthermore, to account for the impact of OD distribution, 100 replications are conducted for all the tests. Each point in the following figures represents the average value of 100 replications and vertical bars for standard deviations.

*Algorithmic Performance*

- Pruning Performance

The pre-processing step utilizes only the time constraints to prune the infeasible passenger requests. That is, restricted excess time constraints are expected to generate fewer candidate passengers (high pruning). We evaluate the change of this step and prune strength by only changing the excess travel time constraints in the default setting.

We define the prune strength as the average percentage of candidate passenger requests for all the ride-sharing drivers:

$$\text{Prune Strength} = \frac{1}{|V|} \sum_{v \in V} 1 - \frac{N_{CR}^v}{|R|} \% \qquad (32)$$

Yao, Bekhorwhere $N_{CR}^v$ represents the total number of candidate passenger requests for driver *v*.

Table 3 Prune time and prune strength with respect to excess time constraints

| **Excess Time Constraint** | 300% | 200% | 100% | 50% | 20% | 10% |
|---|---|---|---|---|---|---|
| **Pre-processing Step Time [ms]** | 8.58 | 8.39 | 6.95 | 5.78 | 5.36 | 5.31 |
| **Prune Strength** | 5.90% | 17.94% | 59.06% | 88.76% | 98.31% | 99.45% |

Overall, the pre-processing step time is negligible, and the effect of different excess travel time constraints is relatively small to the prune time. As expected, the stricter excess travel time constraints are, the stronger is the prune strength. For example, with 20% excess travel time constraint, the average number of candidate passenger requests is about 35.

- Runtime Performance

Runtime performance is composite of Total Runtime, Feasible Request Combinations Generation Time and ILP Solving Time. Note that pre-processing step times are orders of magnitude lower and are neglected here.

We evaluate the impact of driver supply (Figure 9), excess travel time constraints (Figure 10) and vehicle capacities (Figure 11). For each of the evaluations, only the selected parameter is changed while others remain the same as the default setting.

Table 4 Selected parameters - default values and analysis values

| **Parameters** | **Default** | **Analysis Values** |
|---|---|---|
| Driver-to-Passenger Ratio* | 1:2 | 1:5, 1:4, 1:3, 1:2, 1:1, 2:1, 3:1, 4:1, 5:1 |
| Vehicle Capacity | 4 | 1, 2, 3, 4 |
| Excess Time Constraints** | 20% | 10%, 20%, 50%, 100%, 200%, 300% |
| *: total of 3,000 ride-sharing participants, ratios are converted to driver percentages in the following figures ||| 
| **: percentage to the participants' shortest path travel times |||

Yao, Bekhor

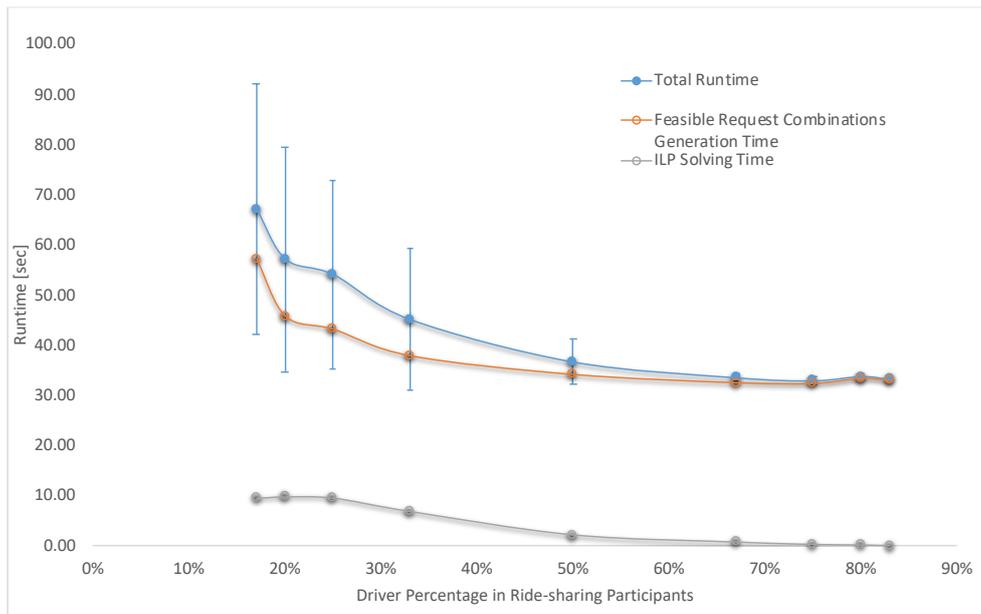

Figure 9 Driver percentage impact on runtime

The impact of driver percentage on runtime is related to the complexity analysis. The complexity of feasible passenger request combinations is of order $O(|V||R|^m)$, which is linear to the number of drivers but to the power of $m$ to the number of passengers. The result is shown in Figure 9, as the driver percentage decreases, the passenger percentage increase while the runtime is increased greater than linear but much better than the worst case (power of 4) due to the application of solution properties in the dynamic tree algorithm.

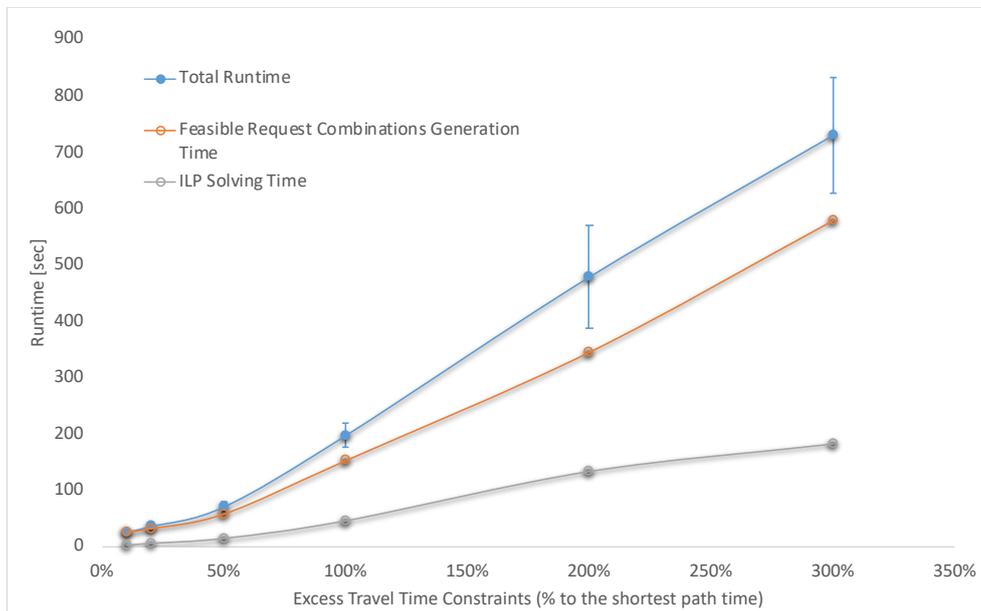

Figure 10 Excess travel time constraints impact on runtime

Figure 10 shows that runtimes are very sensitive to the excess travel time constraints. As shown in Table 3, for less restrictive excess travel time constraints there will be more potential passenger requests. In these cases, the actual complexity increases, because $|R|$ increases.

Yao, Bekhor

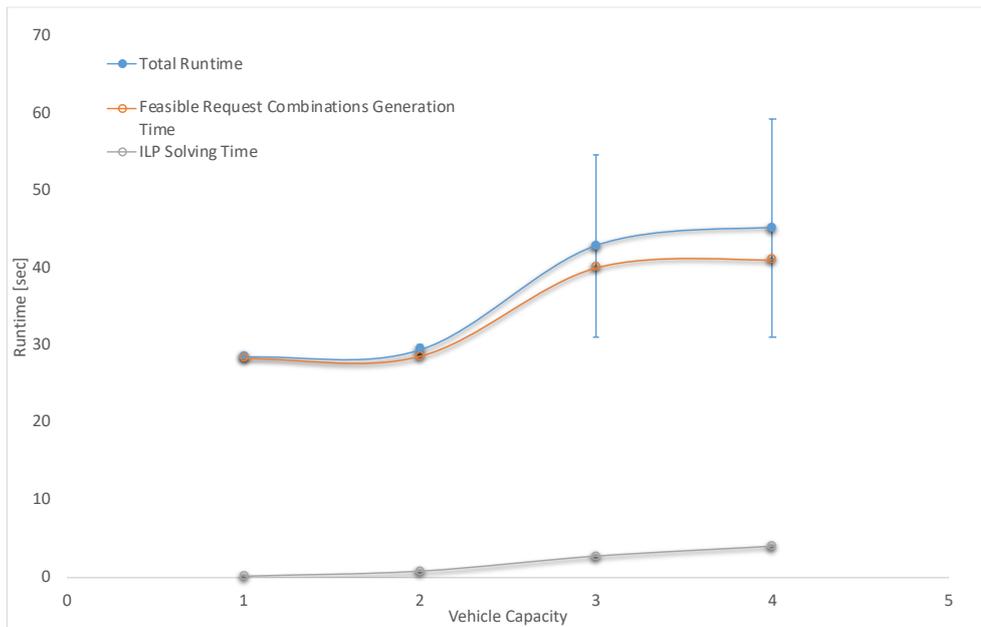

Figure 11 Vehicle capacities impact on runtime

Figure 11 shows that the runtimes are relatively insensitive to the vehicle capacities. The impact of vehicle capacity on runtime is reflected in the dynamic tree algorithm. Larger vehicle capacity means more feasible permutations of passenger requests in which more branches need to be generated and checked.

*Matching Statistics*

The matching results are exemplified in terms of Match Success Rate, Driver and Passenger Actual Excess Travel Time with respect to different driver percentages in ride-sharing participants, different excess travel time constraints, and vehicle capacities. We illustrate the mixed effect of the selected parameters in 3D figures.

- Match Success Rate

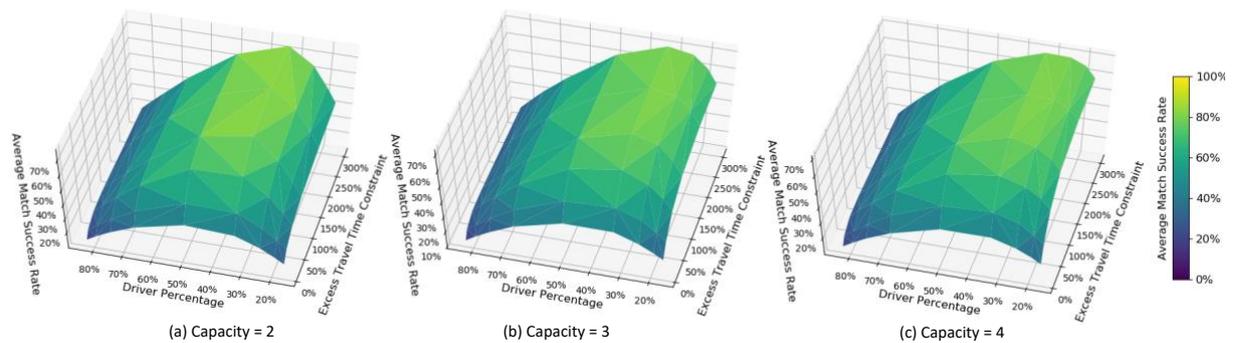

Figure 12 Impact on match success rate

For different vehicle capacities, the match success rates share similar patterns: the maximum and minimum match success rates are similar; match success rates increase with excess travel time constraints; maximum match success rates are observed around 30% – 50% driver percentage and decrease for larger or smaller percentage.

The maximum match success rate occurs at smaller driver percentages for higher capacities as expected, since less vehicles are required to maintain the same level of match success rate with larger capacity vehicles.

- Actual Excess Travel Time

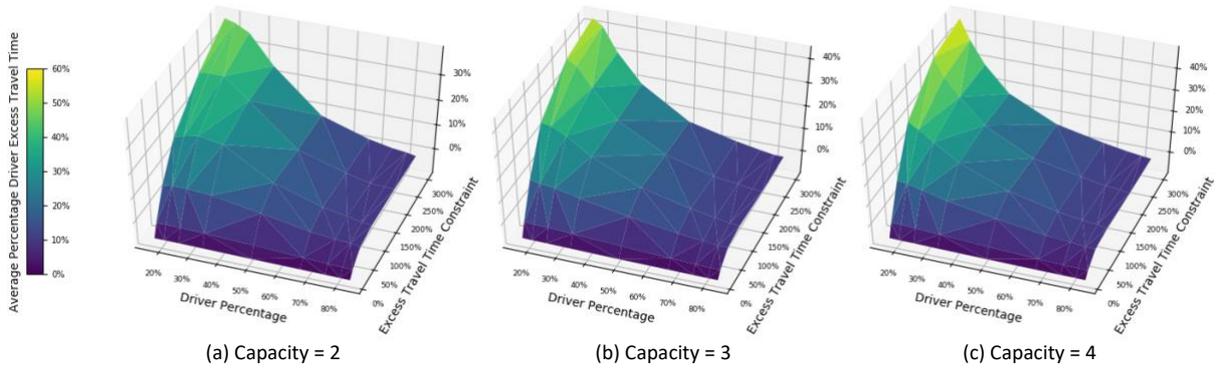

Figure 13 Impact on driver actual excess travel time

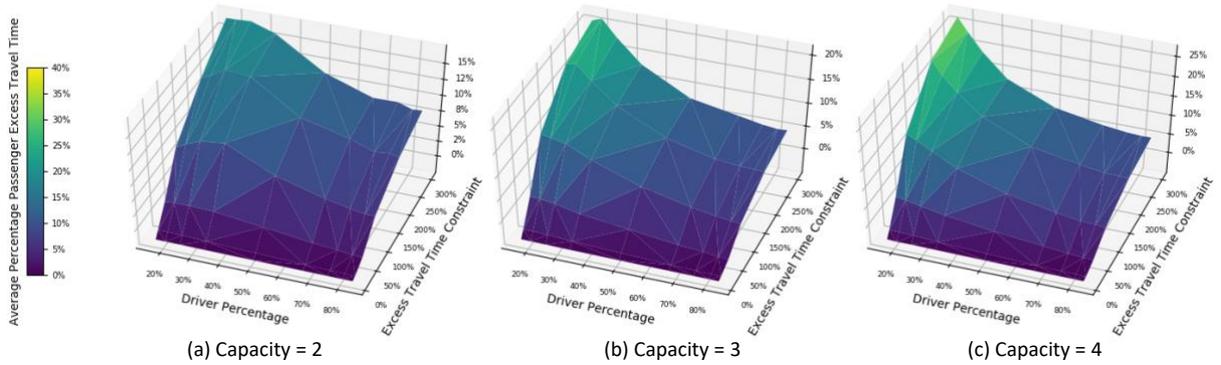

Figure 14 Impact on passenger actual excess travel time

The actual excess travel times, for both drivers and passengers, increase with less restrictive excess travel time constraints, and decrease with larger driver percentages. For larger vehicle capacities, the actual excess travel times also increase. This is also expected, because the driver is able to pick up more passengers and consequently increase his/her detour.

Note that in all cases, passengers' actual excess travel times are smaller than drivers'. This is explained by the objective function, which is to maximize the overall VKT savings. The objective function eq. (4) is composed of two terms: drivers' ride-sharing route travel times, and all participants' (drivers and passengers) shortest path travel times. Therefore, in order to increase the overall VKT savings, the algorithm tries to find the minimum possible actual travel passenger times.

*Ride-sharing Impact on the Network*

We evaluate the impact of the proposed ride-sharing system with different parameters on the transportation network in terms of VKT saved and Vehicle Trips Saved. The original vehicle trips of matched ride-sharing passenger and driver are combined into ride-sharing vehicle trips, i.e. their original vehicle trips are removed from the origin-destination matrix. While majority of the private vehicle trips are drive-alone trips and are remained in the matrix. We also assume that unmatched passengers and drivers are reassigned to the network as drive alone. And a traffic assignment is run to obtain the VKT and VHT. The VKT saved and vehicle trips saved are compared to the base case (without ride-sharing). Figure 15 illustrates the impact of driver percentage, excess travel time constraints, and vehicle capacity.

Yao, Bekhor

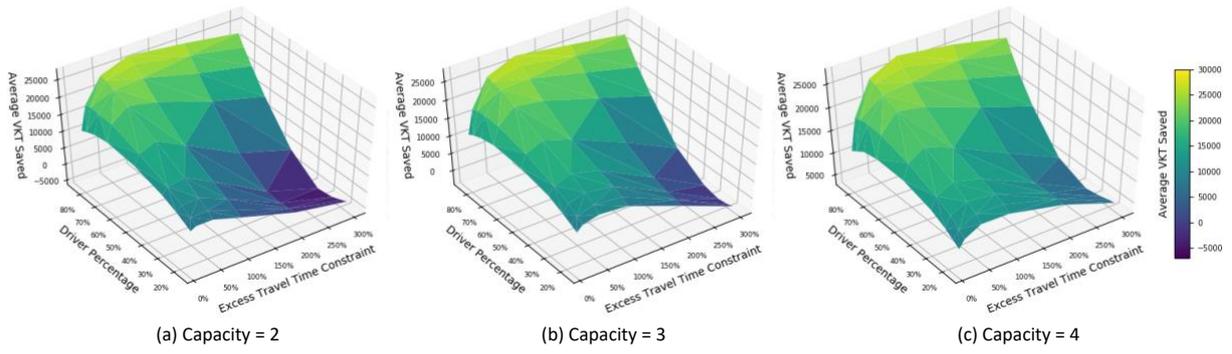

(a) Capacity = 2    (b) Capacity = 3    (c) Capacity = 4

Figure 15 Impact on VKT saved

For all the cases with different capacities, VKT saved reaches the minimum at around 20% driver supply share and 300% excess travel time constraints. Note that negative VKT savings are observed in cases with capacity 2 and 3. Drivers without restrictive excess time constraints (e.g. larger than 200%) take a very long detour to pick up and drop off passengers and bring extra burden to the current network. These results are related to the observations in Clewlow and Mishra (2017) and Erhardt et al., (2019), in which ride-sharing systems may contribute to the growth of VKT in cities. Our tests suggest that the parameters should be carefully selected to maintain the sustainability of a ride-sharing system.

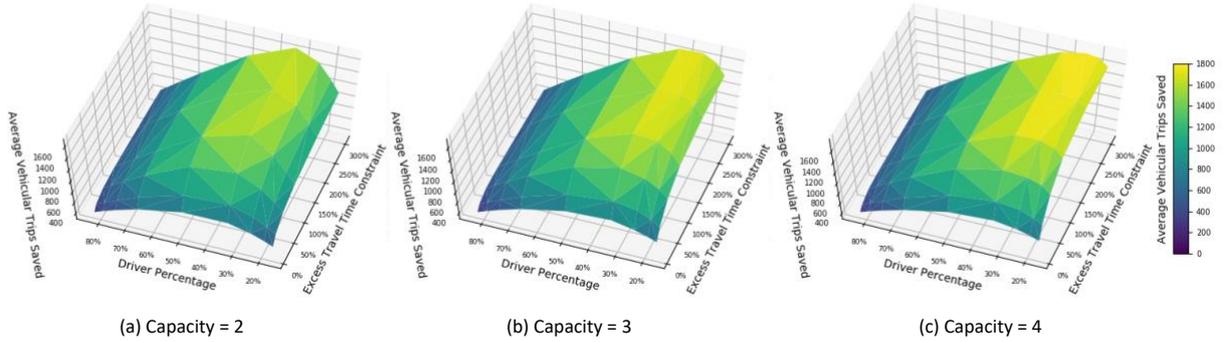

(a) Capacity = 2    (b) Capacity = 3    (c) Capacity = 4

Figure 16 Impact on vehicle trips saved

Vehicle trips saved shown in Figure 16 agrees with the match success rate shown above. Note that Figure 16 shows that high match success rate (or vehicle trips saved) does not guarantee maximum VKT saved.

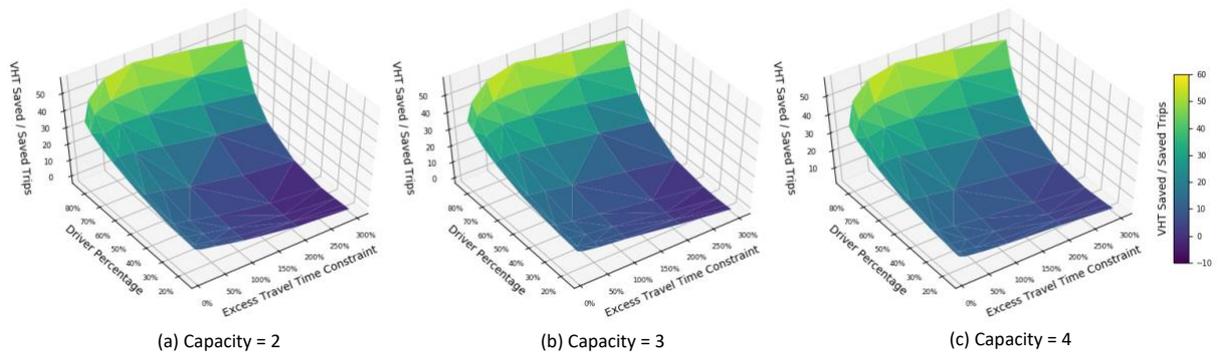

(a) Capacity = 2    (b) Capacity = 3    (c) Capacity = 4

Figure 17 Impact on VKT saved per saved trip

Figure 17 illustrates the average VKT saved per Saved Vehicle Trip. For high driver percentages, and consequently more drivers available to the passengers, the algorithm results in less detours, and thus larger average VKT saved per trip are observed.

**DISCUSSION**

The result section illustrated that our proposed dynamic tree algorithm reaches the same objective function values and smaller runtimes in comparison with the exact solution algorithm for simple problems. The efficiency and effectiveness of the proposed algorithm to solve a real size problem are also examined using a large example.

One of the challenges still remaining in on-demand ride-sharing system implementation is how to efficiently handle a large number of requests with similar origins and destinations within a relatively small area. This is particularly complicated, because of the need to consider matching many potential passenger requests with a driver, and our proposed pruning method will still leave multiple possible combinations. To cope with this issue, we propose the following:

1) Consider stricter pruning rules (i.e. stricter maximum excess travel time constraints than the ones specified by drivers and passengers), and consequently reduce the number of potentially feasible requests. This option potentially may harm the objective function values. However, as presented in the sensitivity analysis section (Figure 15), restricting maximum excess travel time constraint within some range, (e.g. between 150% to 80% of the original value) does not change very much the overall VKT saved.
2) Another idea is to group drivers and passenger requests with the same or similar ODs and constraints, respectively. Similar passenger requests are grouped into one request with multiple passengers and the OD of the grouped request is placed at the center of these requests' origin and destination respectively, similar to the approach adopted in Huang et al. (2017). Moreover, only one set of request combinations needs to be generated for drivers with similar ODs and constraints, and then duplicated for each of these drivers for solving the ILP. These approaches can significantly reduce the computations for passengers and drivers traveling with similar ODs.
3) According to the complexity analysis, the time-consuming steps are Dynamic Tree step with $O\big((2m)!\big)$ and the Request Combination Generation step with $O(|V||R|^m)$. The adjustable parameter $m$, which determines the number of potential passenger requests considered in both steps, has a great impact on runtime performance. With a smaller parameter $m$, the feasible search space is reduced which results in faster computation, but at the cost of worse objective function values. In case there are many requests traveling within a relatively small area, one could set a smaller $m$ to allow handling these requests efficiently, as in Alonso-Mora et al. (2017).

We would like to address that the trade-offs between algorithm runtime performance and solution qualities in terms of objective function values should be carefully considered before applying the above methods.

Another interesting point is related to the requests batching time interval in many-to-many ride-sharing matching. The many-to-many ride-sharing matching approach accumulates passenger requests and drivers for a period of time to perform matching.

With a relatively large batching time interval, a larger number of waiting passengers and drivers can be collected. This is expected to reduce the pickup distance and excess travel distance for both drivers and passengers, because we can have a better chance to match drivers and passengers with close origin-destination and less detours using a larger passenger and driver pool.

However, if the batching time interval is excessively long, passengers may be unsatisfied with long response waiting time, and even abandon their requests. And peer-to-peer ride-sharing drivers who have tight schedules might not be able to wait for the response, and eventually quit the service.

The requests batching time interval should be appropriately specified, by taking into considerations the system objectives (in our case, the total vehicle kilometer traveled) and other performance measurements (such as matching success rate, drivers' and passengers' waiting time for system response, waiting time for pickup, excess travel times and distances). In the recent paper of Yang et al. (2020), the importance of matching time interval specification is also addressed in the context of ride-sourcing.

**SUMMARY AND OUTLOOK**

In this paper, we propose an efficient dynamic tree algorithm with a simple and effective pre-processing step to solve the on-demand peer-to-peer ride-sharing matching problem. Similar to Alonso-Mora et al. (2017), our approach decouples the complicated ride-sharing matching problem into simpler parts and solves the problem separately. One of our unique contributions is to integrate the dynamic tree algorithm for solving ride-sharing VRP. The dynamic tree provides the algorithm with the ability to utilize the previously calculated driver schedules, instead of calculating them from scratch. The numerical tests confirm that our proposed dynamic tree algorithm is computationally efficient. Another contribution of this paper is the inclusion of a simple geometric pruning procedure to reduce the size of the problem. Consequently, runtime results for the Pre-processing step are very small, and this procedure helps reduce the overall algorithm runtimes.

This paper exemplifies the algorithm implementation using the well-known Winnipeg network. We performed several tests related to OD distributions, driver supply, excess travel time constraints, and vehicle capacities. Numerical experiments show that the spatial distribution of ride-sharing participants influences the algorithm performance. Sensitivity analysis confirms that the most critical ride-sharing matching constraints are the excess travel times. For the range of vehicle capacities tested (between 1 and 4), the algorithm performs well, and runtime performances are similar.

The network analysis using traffic assignment suggests that ride-sharing systems should be carefully designed to maintain the sustainability of a ride-sharing system, and help alleviate the traffic congestion in the network. In particular, the results suggest that small vehicle capacities do not guarantee overall vehicle-kilometer travel savings.

Moreover, our matching statistics show that passengers' actual excess travel times are smaller than drivers' in general under our settings. This is related to our assumption of the objective function, which is to maximize overall VKT savings. Clearly, incentives should be included to compensate the excess travel times endured by the ride-sharing drivers and passengers.

This paper assumed several parameters related to the demand-side constraints, such as maximum excess travel times, maximum waiting times, and earliest pickup times. The matching results are of course dependent on these parameters. Further research is needed to provide recommendations with respect to ride-sharing level of service. In addition, this paper did not consider advanced booking option, but we believe it is possible to accommodate this option in the proposed solution approach, and will be investigated in a future study.

The excess travel times were calculated by using a fixed travel time matrix. The re-routing of drivers may change the travel times on the network (because of the change in traffic flows), and consequently may change the matching results. Further research will investigate the stability of the matching results with respect to travel times.

**APPENDIX**

*NP-Hardness*

Similar to NP-hardness of DARP (Baugh et al., 1998), we will prove our formulation of on-demand peer-to-peer ride-sharing matching problem is also NP-hard by reducing the known NP-complete Hamiltonian cycle problem (Garey and Johnson, 1978; Baugh et al., 1998) to this problem. Given a graph $G'(N', A')$ to a Hamiltonian cycle problem instance, it tries to find a circle in $G'$ that passes through every node in $G'$ exactly once.

**Theorem 1.** *The proposed on-demand peer-to-peer ride-sharing matching problem is NP-hard*

*Proof:* The decision version of the proposed on-demand peer-to-peer ride-sharing matching problem can be stated as: Given the passenger-driver network $G(N, A)$, set of drivers $V$, set of passenger requests $R$, capacity and time constraints, verify if there is a feasible solution for $z_1$ with an objective value smaller or equal to $Z_1$ and verify if there is a feasible solution for $z_2$ with an objective value larger or equal to $Z_2$.

The decision version of the proposed on-demand peer-to-peer ride-sharing matching problem is shown to be NP-complete by showing that:

- The proposed on-demand peer-to-peer ride-sharing matching problem is in NP:

For given $G, V, R$, set of constraints, and an assignment in which passenger requests are assigned to drivers, there exists a polynomial-time algorithm to check the validity of the given assignment and to verify if the objective value exceeds $Z$ or not.

- The Hamiltonian cycle problem can be reduced to the proposed on-demand peer-to-peer ride-sharing matching problem:

Suppose a graph $G'$ is given to the Hamiltonian cycle problem, we construct a passenger-driver network $G$ from $G'$ and give it to the proposed on-demand peer-to-peer ride-sharing matching problem:

1. Select any node $s$ in $N'$, create an origin node $o_i$ and a destination node $d_i$ for driver $v_i$ in passenger-driver network $G$, let $t_{ED}(v_i) = 0$, $c_i = 1$ and $\Delta_{v_i} = |N'|$. For each node $j \in (N' \backslash s)$, create an origin node $o_j$, a destination node $d_j$, and a request $r_j$, which $t_{ED}(r_j) = 0$, $\Omega_{r_j} = |N'|$ and $\Delta_{r_j} = |N'|$.

2. For every arc $(a, b)$ in $G'$, add an arc $(d_a, o_a)$ of $tt_{(d_a, o_a)} = 1$ and $l_{(d_a, o_a)} = 1$. Also add arcs of $(o_c, d_c)$ for $\forall c \in N'$ with $tt_{(o_c, d_c)} = 0$ and $l_{(o_c, d_c)} = 0$.

We can see that the transformation from $G'$ to $G$ can be performed in polynomial time. If we set $Z_1 = \frac{|N'|(1+|N'|)}{2}$ and penalty factors $M_1 \geq |N'|, M_2 \geq |N'|$ for objective $z_1$, set $Z_2 = |N'|$ for objective $z_2$, both objectives will match all passenger requests, since each node in $G$ can be visited at most once (eq. (9)) and there are bounds $Z$ on objective values. Let $S$ be the route of ride-sharing driver which serves all passenger requests, i.e. passes through each node exactly once, then there is a Hamiltonian cycle in $G'$.

Let $H$ be a Hamiltonian cycle, since all nodes in $N'$ are visited by $H$, so all the requests will also be served, there is a feasible solution for $z_1$ and $z_2$ respectively. ∎